\documentclass[twocolumn]{aastex62}
\usepackage {graphicx}
\usepackage[caption=false]{subfig}
\usepackage{amsmath,amsfonts,amsthm,bm}

\usepackage{hyperref}
\usepackage{xcolor}
\usepackage{verbatim}
\usepackage{booktabs}
\definecolor{medium-blue}{rgb}{0,0,1}
\hypersetup{colorlinks, urlcolor={medium-blue}}

\definecolor{my_color}{HTML}{3a18b1}

\definecolor{new_color}{HTML}{CF0000}% this is a maroon
\definecolor{new_black}{HTML}{000000}% this is a maroon
\usepackage{xfrac}

\newcommand\update[1]{{#1}}

\graphicspath{{figures/}}

\newcommand{\TESS}{\textit{TESS}}

\shorttitle{Improved TESS Triage with Neural Networks}
\shortauthors{Tey/Moldovan \textit{et al.} }
\begin{document}

% REU: The title is the single most important element of the paper, because it is the part
% most widely used for literature searches.  
%
% Your title should be about  WHAT you did, and not tightly focused on HOW you did it.  
% The methods you used are of secondary importance to what you accomplished.  
% In other words, put your science first.  Try not to use the overused word "study" in the title.

\title{Identifying Exoplanets with Deep Learning.~V.~Improved Light Curve Classification for TESS Full Frame Image Observations}

\author[0000-0002-5308-8603]{Evan Tey$^*$}
\affiliation{Department of Physics and Kavli Institute for Astrophysics and Space Research, Massachusetts Institute of Technology, 77 Massachusetts Ave, Cambridge, MA, 02139, USA}
\altaffiliation{These authors contributed equally to the manuscript.}

\author[0000-0003-2553-2081]{Dan Moldovan$^*$}
\affiliation{Google }
\altaffiliation{These authors contributed equally to the manuscript.}

\author[0000-0001-9269-8060]{Michelle Kunimoto}
\affiliation{Department of Physics and Kavli Institute for Astrophysics and Space Research, Massachusetts Institute of Technology, 77 Massachusetts Ave, Cambridge, MA, 02139, USA}

\author[0000-0003-0918-7484]{Chelsea X.  Huang}
\affiliation{University of Southern Queensland, Centre for Astrophysics, West Street, Toowoomba, QLD 4350 Australia}

\author[0000-0002-1836-3120]{Avi Shporer}
\affiliation{Department of Physics and Kavli Institute for Astrophysics and Space Research, Massachusetts Institute of Technology, 77 Massachusetts Ave, Cambridge, MA, 02139, USA}

\author[0000-0002-6939-9211]{Tansu Daylan}
\affiliation{Department of Physics and Kavli Institute for Astrophysics and Space Research, Massachusetts Institute of Technology, 77 Massachusetts Ave, Cambridge, MA, 02139, USA}
\affiliation{Department of Astrophysical Sciences, Princeton University, 4 Ivy Lane, Princeton, NJ 08544}
\affiliation{LSSTC Catalyst Fellow}

\author[0000-0002-5788-9280]{Daniel Muthukrishna}
\affiliation{Department of Physics and Kavli Institute for Astrophysics and Space Research, Massachusetts Institute of Technology, 77 Massachusetts Ave, Cambridge, MA, 02139, USA}

\author[0000-0001-7246-5438]{Andrew Vanderburg}
\affiliation{Department of Physics and Kavli Institute for Astrophysics and Space Research, Massachusetts Institute of Technology, 77 Massachusetts Ave, Cambridge, MA, 02139, USA}

\author[0000-0002-1092-2995]{Anne Dattilo}
\affiliation{Department of Astronomy and Astrophysics, University of California, Santa Cruz, CA 95064, USA}

% TODO: architects

\author[0000-0003-2058-6662]{George~R. ~Ricker}
\affiliation{Department of Physics and Kavli Institute for Astrophysics and Space Science, Massachusetts Institute of Technology, 77 Massachusetts Ave, Cambridge, MA, 02139, USA}

\author[0000-0002-6892-6948]{S.~Seager}
\affiliation{Department of Physics and Kavli Institute for Astrophysics and Space Science, Massachusetts Institute of Technology, 77 Massachusetts Ave, Cambridge, MA, 02139, USA}
\affiliation{Department of Earth, Atmospheric and Planetary Sciences, Massachusetts Institute of Technology, 77 Massachusetts Ave, Cambridge, MA, 02139, USA}
\affiliation{Department of Aeronautics and Astronautics, Massachusetts Institute of Technology, 77 Massachusetts Avenue, Cambridge, MA 02139, USA}

\begin{abstract}
The TESS mission produces a large amount of time series data, only a small fraction of which contain detectable exoplanetary transit signals. Deep learning techniques such as neural networks have proved effective at differentiating promising astrophysical eclipsing candidates from other phenomena such as stellar variability and systematic instrumental effects in an efficient, unbiased and sustainable manner. This paper presents a high quality dataset \update{containing light curves} from the Primary Mission and 1st Extended Mission full frame images \update{and periodic signals detected via Box Least Squares \citep{2002A&A...391..369K, 2012ascl.soft08016H}}. The dataset was curated using a thorough manual review process then used to train a neural network called \texttt{Astronet-Triage-v2}. On our test set, for transiting/eclipsing events we achieve a 99.6\% recall \update{(true positives over all data with positive labels)} at a precision of 75.7\% \update{(true positives over all predicted positives)}. Since 90\% of our training data is from the Primary Mission, we also test our ability to generalize on held-out 1st Extended Mission data. Here, we find an area under the precision-recall curve of 0.965, a 4\% improvement over \texttt{Astronet-Triage} \citep{2019AJ....158...25Y}. On the TESS Object of Interest (TOI) Catalog through April 2022, a shortlist of planets and planet candidates, \texttt{Astronet-Triage-v2} is able to recover 3577 out of 4140 TOIs, while \texttt{Astronet-Triage} only recovers 3349 targets at an equal level of precision. In other words, upgrading to \texttt{Astronet-Triage-v2} helps save at least 200 planet candidates from being lost. The new model is currently used for planet candidate triage in the Quick-Look Pipeline \citep{2020RNAAS...4..204H, 2020RNAAS...4..206H,Kunimoto2021}.
\end{abstract}

% https://astrothesaurus.org/concept-select/

\section{Introduction}
\label{sec:intro}

For three decades, human judgement has played a critical role in the exoplanet revolution that has yielded the discovery of more than 5000 planets outside of the Solar System\footnote{NASA Exoplanet Archive: \href{exoplanetarchive.ipac.caltech.edu}{exoplanetarchive.ipac.caltech.edu}}. Exoplanets are typically much cooler, smaller, and fainter than their host stars, so detecting them usually requires extremely precise observations. At the level of sensitivity required to detect exoplanets, numerous other systematic effects can be present in data that can mimic planetary signals. Separating out these ``false positive'' signals from true exoplanets has been a major challenge \citep{jacob, vandekamp, bailes} since before the discovery of the first exoplanets in the 1980s and 1990s \citep{campbellwalker, latham1989, Wolszczan, mayor}. Historically, classifying possible planet signals as either false positives or viable planet candidates has most often been carried out by a human inspecting and making a judgement on each signal. Humans are quite well suited for this type of work; we can learn how to distinguish planet candidates and false positives with high accuracy, even after looking at a relatively small number of examples, and often without the benefit of \textit{a priori} knowledge of the ``ground truth'' of any signal's true classification. 

However, relying on human judgement to separate viable planet candidates from false positives has two main disadvantages. First, humans are slow, both in terms of training time and actual classifications. It often takes months or years of practice for a human to become adept at classifying planets and false positives, and once fully trained, it may take an experienced human several minutes to review all of the information needed to make one classification. At these speeds, even classifying a modest number of possible planet signals ($\sim 10^2-10^3$) may take days. Given the rapid increase in the volume of astronomical data available for analysis, it will soon be impractical to rely on human classifications to identify viable planet candidates. Second, humans are inconsistent. Differences in external factors (mood, fatigue, hunger, etc) may cause a human to judge the same signal differently on two different occasions. This makes characterizing and quantifying the biases introduced by human classification challenging and inexact. An alternative system capable of quickly, accurately, and repeatably identifying planet candidates would be highly attractive to planet hunters.

In this paper, we focus on improving a deep neural network classifier used to identify viable planet candidates in data from the \textit{Transiting Exoplanet Survey Satellite} (\textit{TESS}) mission \citep{ricker15}.  \TESS\ identifies exoplanets by searching for ``transits,'' or slight periodic dimmings of the apparent brightness of a star as its planet passes between the star and our vantage point in the Solar System. Transit surveys like \TESS\ produce copious numbers ($\gtrsim 10^6$ so far) of false positive signals that must be separated from viable planet candidates to enable discoveries.  

Machine learning has become a popular tool for identifying promising planet candidates from transiting exoplanets. Some work has focused on using machine learning to perform the actual planet detection \citep{pearson, zucker, cui}, but more often, efforts have focused on using machine learning to classify the large number of possible transit-like signals returned by existing planet detection pipelines. A push early in the Kepler mission \citep{kochkepler, borucki2010} led to the development of two automated systems: a decision tree called the Robovetter \citep{coughlincatalog,thompsoncatalog} and a random forest classifier called the Autovetter \citep{autovetter}. In that initial work, the Robovetter proved more robust and easily extensible to new regimes and datasets, and therefore was used in the production of fully automated planet candidate catalogs from the Kepler mission. 

More recently, \citet{2018AJ....155...94S} introduced a convolutional neural network for vetting planet candidates from the Kepler mission called \texttt{Astronet}. Since then, \texttt{Astronet} and other similar architectures have been demonstrated on other datasets like K2 \citep{2019AJ....157..169D}, TESS \citep{2019AJ....158...25Y, 2020A&A...633A..53O}, WASP \citep{2019MNRAS.483.5534S}, and NGTS \citep{2018MNRAS.478.4225A, 2019MNRAS.488.5232C}. New tweaks to the methdology including new input information and tweaks to the data representation \citep{2018ApJ...869L...7A,jara2020transiting, 2021arXiv211110009V} have yielded improvements in classification performance. 

Our work is largely based upon the convolutional neural network originally introduced by \citet{2018AJ....155...94S} and adapted to \TESS\ by \citet{2019AJ....158...25Y}, known as \texttt{Astronet-Triage}. Starting in 2019, \texttt{Astronet-Triage} had been used in the \TESS\ Quick-Look Pipeline \citep{guerrero} to triage planet candidates and remove clear false positives. However, our internal tests revealed that this step resulted in the loss of a fairly large number of viable planet candidates (i.e., ``false negatives''). This paper describes our work to improve the performance of \texttt{Astronet-Triage} by introducing \texttt{Astronet-Triage-v2} to reduce the number of lost planet candidates while throwing out a higher number of false positives. 

Our paper is organized as follows: In Section \ref{observation_descrip}, we describe the input transit signals and corresponding light curves which were used for training and testing our classifier, and the labels assigned to each signal. In Section \ref{inputrepresentation}, we describe how we processed the data before it is input to our neural network classifier. In Section \ref{architecture}, we describe the architecture of the neural network and the training process. We quantify and present the results of our classifier in Section \ref{results}, and we discuss the implications of these results in Section \ref{discussion}. Finally, we conclude in Section \ref{conclusions}.

\section{Data}\label{observation_descrip}
For training and testing our model, we use approximately 25000 human vetted transit signals detected by the Quick-Look Pipeline \citep[QLP,][]{2020RNAAS...4..204H, 2020RNAAS...4..206H,Kunimoto2021} across Sectors 1 -- 39.\footnote{QLP data can be found at \dataset[doi:10.17909/t9-r086-e880]{https://dx.doi.org/10.17909/t9-r086-e880}}

\subsection{TCEs from TESS FFIs}\label{tess_data}
During its Prime Mission (2018 July 25 -- 2020 July 04), TESS collected full-frame images (FFIs) every 30 minutes for 2 years covering 70\% of the entire sky \citep{guerrero}. The FFI cadence was updated to 10 minutes for the 1st Extended Mission (2020 July 04 -- 2022 September 01). QLP produces light curves from these images for all observed targets in the TESS Input Catalog \citep[TIC;][]{2018AJ....156..102S, 2019AJ....158..138S, Paegert2021} with TESS-band magnitude ($T$) brighter than 13.5. 
Flux time series (raw light curves) from five different sized circular apertures are extracted for each star. 

These raw light curves are then filtered to remove low-frequency variability originating from stellar activity or instrument noise. Primarily, this is done by dividing the light curve from each separate orbit by a basis spline  \citep[following ][]{2014PASP..126..948V} fit using a break-point spacing between 0.3 days and 1.5 days, selected as described by \citet{2018AJ....155...94S}. Finally, these detrended light curves are merged with previous TESS sectors using a shared median value. At this point, an optimal aperture is selected for target star based on its TESS magnitude -- fainter stars getting smaller aperture sizes. All subsequent processes use these multi-sector ``best''-aperture detrended light curves.

QLP searches these light curves for transit signals using the Box Least Squares (BLS) algorithm \citep{2002A&A...391..369K, 2012ascl.soft08016H}. Because BLS spectra feature a rising trend towards lower frequencies (longer periods), QLP subtracts the low frequency baseline before selecting the highest peak as the detection. For each detected signal, the BLS implementation computes characteristic parameters (orbital period, transit center, transit depth, the full transit duration) by performing a least square trapezoid fit for the transit. These parameters are used later in the input process for \texttt{Astronet-Triage-v2}.

Transit signals with signal-to-pink-noise $>9$ and BLS peak significance $>5$ (for stars with $T < 12$ mag) or $>9$ (for stars with $T > 12$ mag) are labelled threshold-crossing events (TCEs). These filters give slightly different perspectives on transit significance: (1) signal-to-pink-noise compares the transit depth to pink noise in the light curve \citep{Pont2006}, while (2) BLS peak significance compares the BLS spectrum's peak height to its noise. In combination, these checks help filter out events that are clearly not transit-like.

In addition, we filter out instances where the planet would orbit ``inside the star.'' For each signal we compute the expected semi-major axis to stellar radius ratio assuming a Keplerian orbit.\footnote{When computing the semi-major axis we use two times the detected BLS period in case the detected period is half the true period, which often happens for eclipsing binaries. If the star has an estimate for its mass in the TIC, we use that value; if not, we assume a mass of 1 $M_\odot$. We also assume a circular orbit. } If the ratio $<1$, the signal is labeled as inside the star. Typically, these signals signify stellar variability or blended signals from a smaller nearby star.

\subsection{Assembling a set of signals to label}

Even with filters described in the previous subsection, manually labeling every TCE would take an enormous amount of time, so we select a subset of TCEs for training / testing. Over time, we gradually accumulated three batches of labeled TCEs from the first two years of TESS Primary Mission (observed with 30 min cadence) and the first year of the TESS 1st Extended Mission (observed with 10 min cadence).

The year 1 (Y1) TESS observations for the southern hemisphere went through significant changes in noise property due to the spacecraft pointing strategy change in Sector 4,\footnote{\url{https://archive.stsci.edu/missions/tess/doc/tess_drn/tess_sector_04_drn05_v04.pdf}} and the subsequent tweaking of the momentum dump frequency. We selected 8992 TCEs detected in Sector 13 (the last sector of Y1) for the labeling. This was not an intentional choice, but after spending hundreds of person-hours labeling these TCEs, we opted to make use of them regardless. Fortunately, despite the fact that our Y1 TCEs came only from Sector 13, the observations that led to these detections still included a diversity of spacecraft pointing control strategies and data artifacts (for example detector warmups following instrument anomaly events\footnote{\url{https://archive.stsci.edu/missions/tess/doc/tess_drn/tess_sector_08_drn10_v02.pdf}}). In particular, stars observed in Sector 13 have been observed by TESS in Y1 between one to thirteen sectors and cover a variety of prior sectors.

For the year 2 (Y2) TESS observations in the northern hemisphere, the data has more uniform characteristics including a consistent momentum dump frequency of every 4.4 days starting in Sector 14\footnote{\url{https://archive.stsci.edu/missions/tess/doc/tess_drn/tess_sector_14_drn19_v02.pdf}}. We sorted TCEs by their target's TESS magnitude, and then took the 13372 brightest TCEs detected from Sectors 14--26.

In year 3 (Y3), TESS returned to observe the southern hemisphere, with faster cadence and a further improved momentum dump strategy (only once each orbit)\footnote{\url{https://archive.stsci.edu/missions/tess/doc/tess_drn/tess_sector_27_drn38_v02.pdf}}. We added an additional 2588 TCEs from Sectors 27-39, which increased the sky coverage and brightness range for our southern hemisphere labels.

We note that TCEs around stars only observed in one of the CCDs in Sector 13 Camera 1, and Camera 1 and 2 for Sector 24 and 25 are not included in our sample due to temporary unavailability of the data at the time of vetting.

\begin{figure*}
    \centering
    \includegraphics[width=\textwidth]{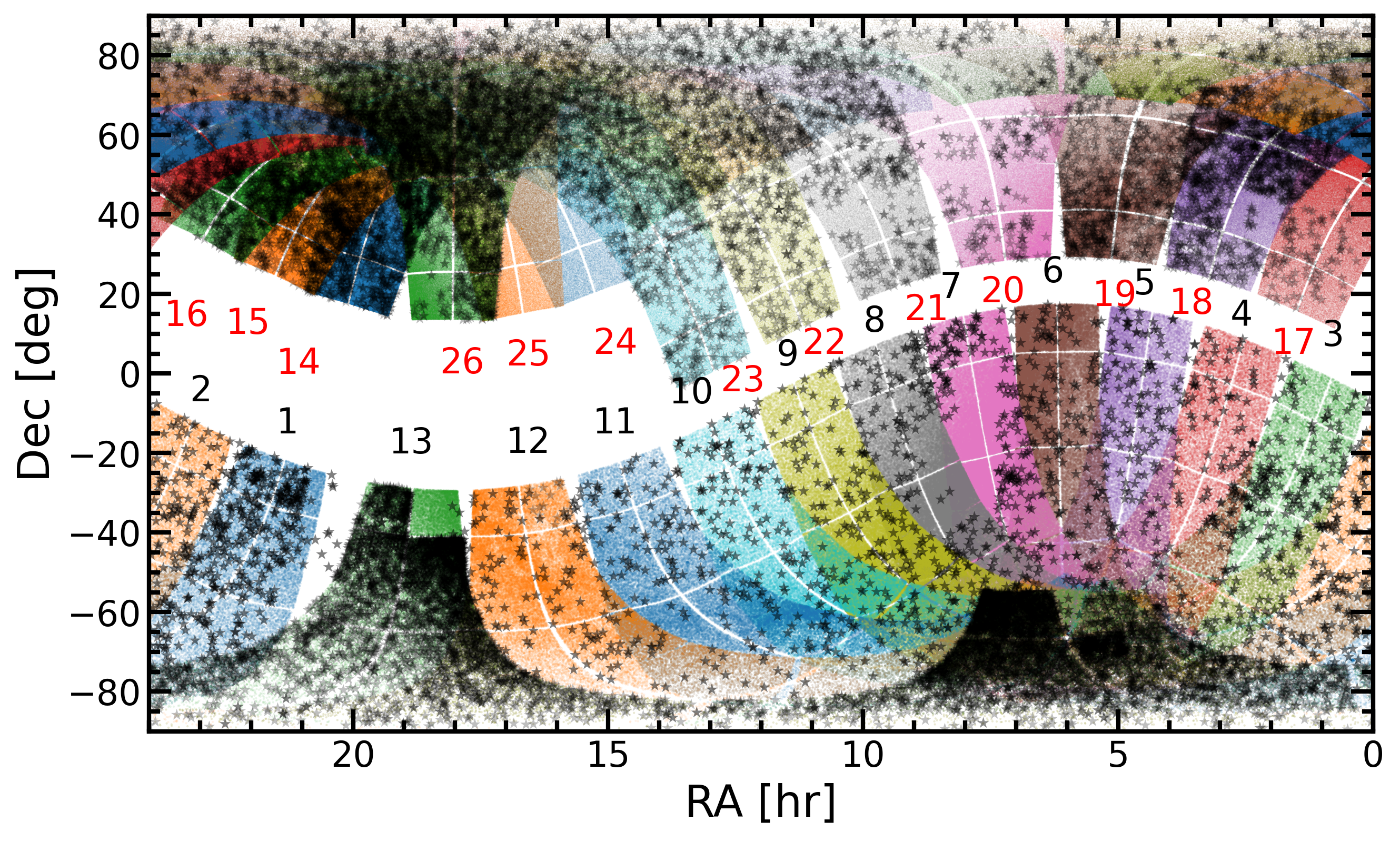}
    \caption{Sky map showing the locations of the 24926 TCEs presented here (black starred data points) compared to the coverage of each TESS Prime Mission sector (colored data points). The black and red labels are the Prime Mission sector numbers in the southern and northern ecliptic hemispheres, respectively. Note that we also include 2588 TCEs from the 1st Extended Mission, for which sector coverage is not shown here. The under- and over-densities of TCEs are due to the selection criteria as described in the text.}
    \label{fig:skymap}
\end{figure*}

\begin{figure}
    \centering
    \includegraphics{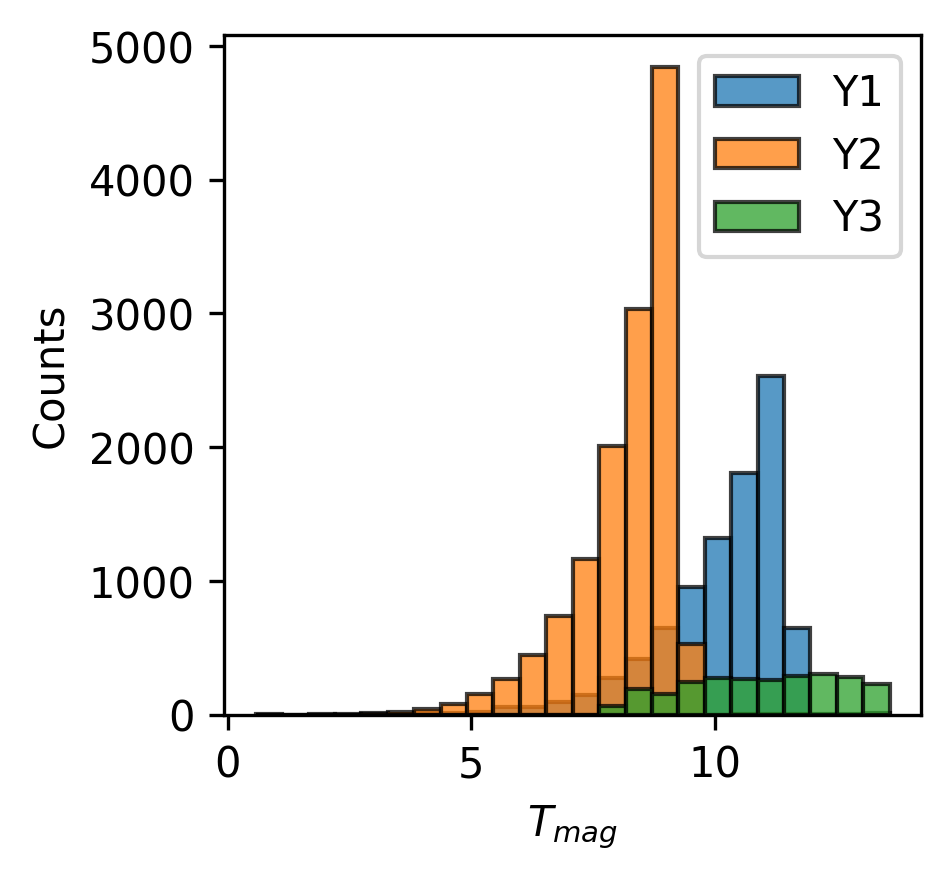}
    \caption{Distribution of $T_{\rm mag}$ across our dataset. Both Y1 and Y2 portions of the dataset focused on the brightest TCEs, while Y3 added TCEs more uniformly across magnitudes. More details on TCE selection can be found in Section \ref{tess_data}.}
    \label{fig:tmag_distr}
\end{figure}

Altogether, these TCEs create a broad sample of transit-like events detected in the first three years of TESS observation. The final TCE distribution across the sky is shown in Figure~\ref{fig:skymap}, and across \TESS\ magnitude ($T_{\rm mag}$) in Figure~\ref{fig:tmag_distr}. Due to the different selection criteria of the TCEs from three different years, they have somewhat different data characteristics. As discussed in Section \S \ref{sec:em_performance}, these differences do not significantly impact our results.  

\subsection{Labels and their definitions}\label{labels}
For each TCE we assigned one of the following five labels:
\begin{itemize}
    \item \textbf{E} denotes a \textit{periodic eclipsing signal}. This includes both planetary transits and non-contact eclipsing binaries. In the triage process, we do not take into account information that would distinguish an eclipsing signal from background stars from an eclipsing signal on the target star. Both cases would be labeled as E if they satisfy all the other criteria. 
    \item \textbf{S} denotes events containing only a \textit{single transit} or events where an \textit{incorrect period} or \textit{period alias} is assessed to be reported from BLS.
    \item \textbf{B} denotes \textit{contact eclipsing binaries}. They are distinguishable from non-contact binaries through their continuous ingress/egress slope.
    \item \textbf{J} denotes \textit{junk}. This includes other astrophysical phenomena like stellar variability as well as instrumental phenomena like scattered light (due to the Earth or the Moon approaching the field of view and reflecting light into the camera) or artifacts introduced at the times of spacecraft momentum dumps (when the spacecraft's reaction wheels correct for the spacecraft's speed).
    \item \textbf{N} denotes \textit{not sure}. No conclusive label decision could be made for these TCEs. Often an N label was given when a weak signal bordered on being an E or J.
    
\end{itemize} 
  
 These labels are not necessarily mutually exclusive. We detail the rules we use in labeling when resolving marginal/ambiguous cases:
 \begin{itemize}
    \item E vs S: If there is ambiguity in the period (e.g.~both the reported period and the double period are consistent with the data) or the period is only slightly off, we default to an E label. Only if the period is explicitly incorrect (e.g there are flat light curve segments during expected transits, or there are multiple regular transits outside of expected transit times) do we choose an S label. If there is only one regular transit outside the expected transit time, i.e. it might represent a secondary eclipse, we use an E label, and if the reported period potentially includes the secondary eclipse, we also use an E label.
    \item B vs S: If we have a contact binary with the incorrect period, we default to a B label.
\end{itemize}

We choose these labels first because they mirror astrophysical phenomena. This means the labeled TCEs provide good targets for follow-up (e.g.~Es will be good candidates for exoplanet and binary star detection). Second, we expect similarities in light curve morphology within a label. This should help our model learn labels more accurately.

For the purposes of finding exoplanets, we are particularly interested in high precision and recall metrics for E labels. S and N labels may also be important candidates for further investigation.

\subsection{Labeling process}\label{label_practice}

\begin{figure*}[p]
    \centering
    \includegraphics[width=.8\textwidth]{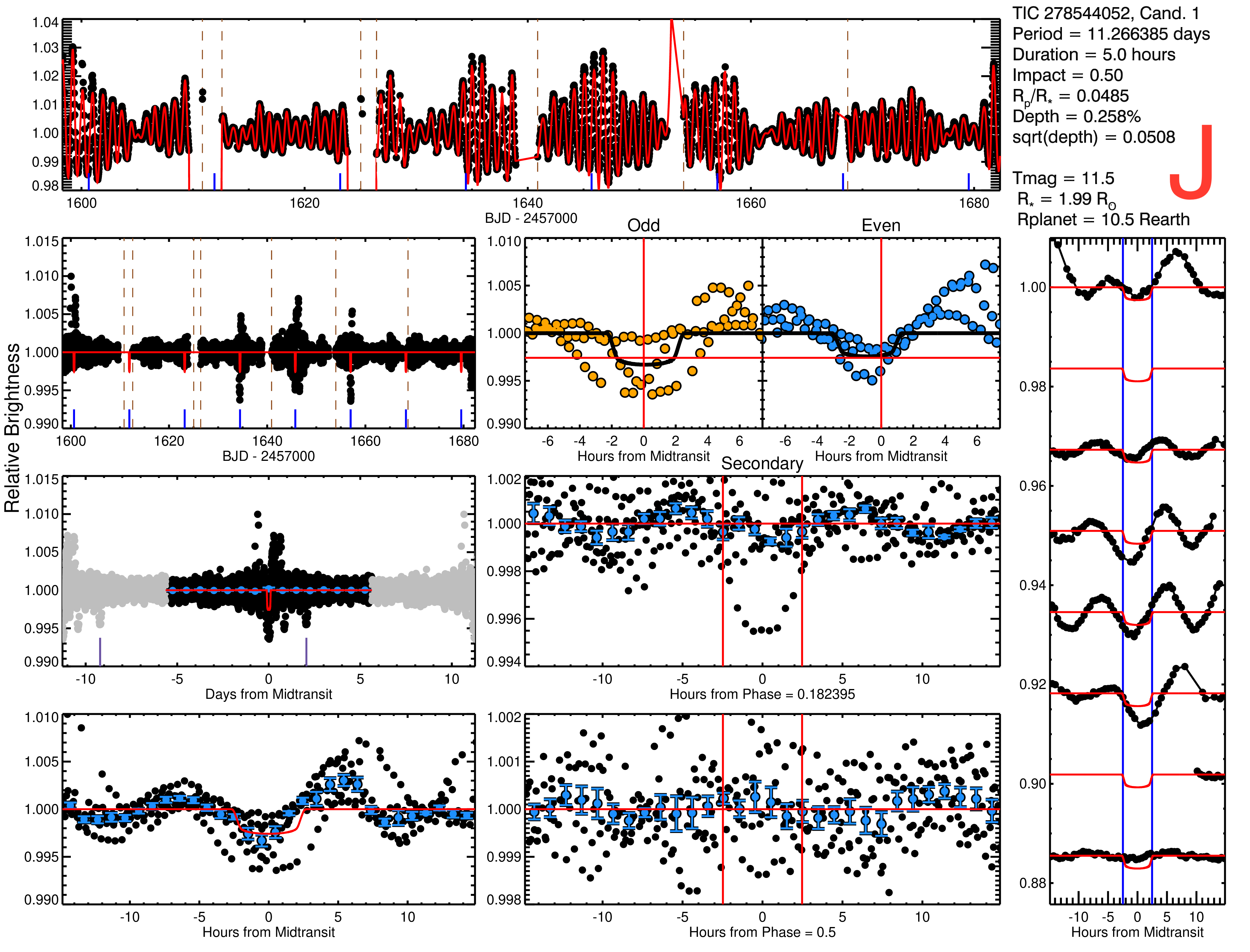}
    \includegraphics[width=.8\textwidth]{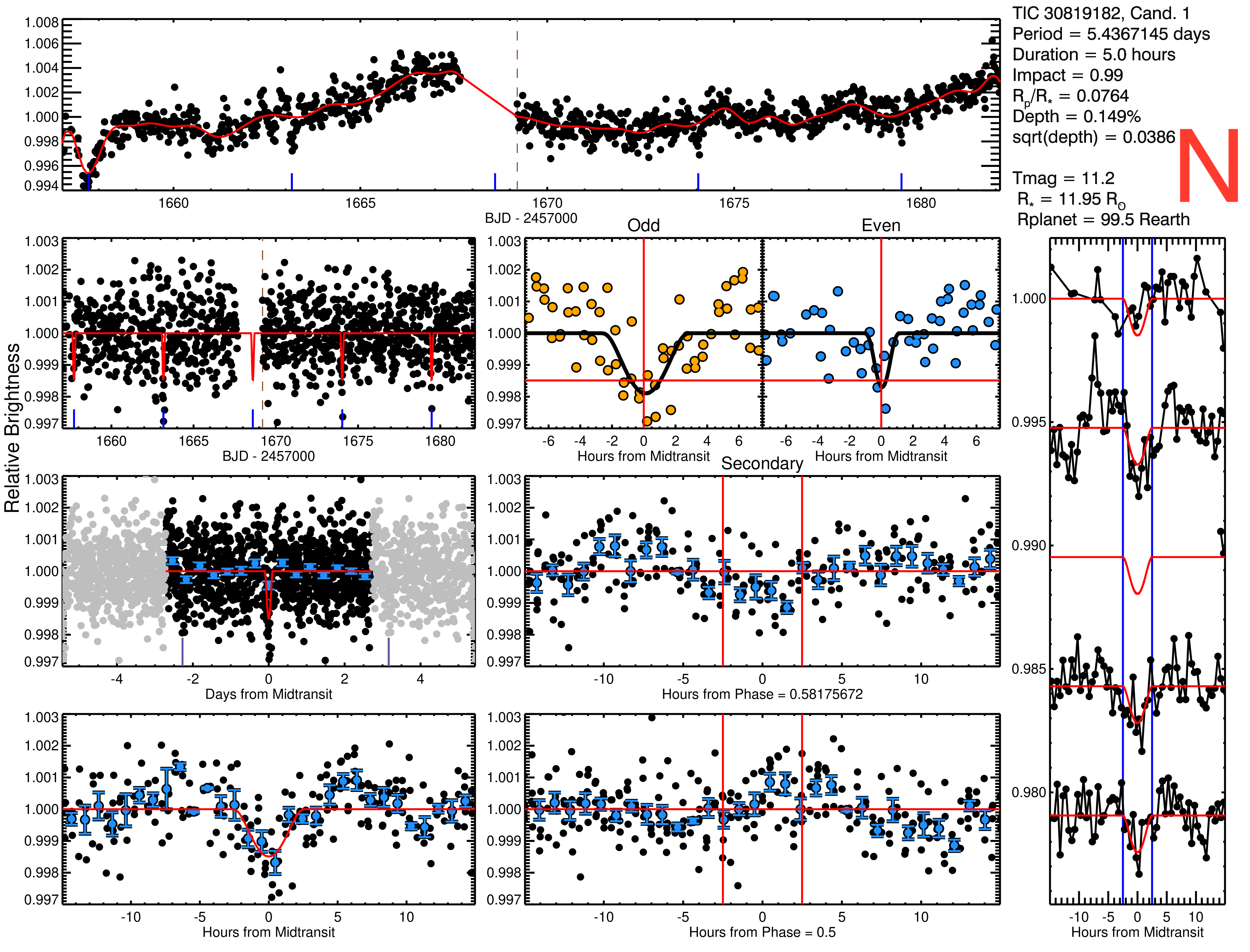}
    \caption{Six example visual representations used for human labeling with labels in red. The different figures within each representation were made to mirror the information described in Section \ref{inputrepresentation}. Each image was individually labeled by at least 3 individual vetters. Conflicting labels were discussed and resolved each week.}
    \label{fig:label_plots}
\end{figure*}

All TCEs were manually assigned labels based on human-visual representations (see Figure \ref{fig:label_plots}) similar to the model input representations described in Section \ref{inputrepresentation}. On a weekly basis, batches of targets were independently vetted by 3 -- 7 of the authors. At the end of the week, targets with conflicting labels where at least one human chose an E or S were discussed in order to reach a consensus on the target's final label. If a target had only B, J, or N votes, we assigned weights to each label based on the number of votes. Altogether, this process took over 2 years. We expect the multiplicity of vetters to reduce the number of label errors, giving us a very high-quality dataset. 

Table \ref{tab:tce_table} contains examples of signal data along with individually-assigned labels and their consensus dispositions. The full table (and accompanying light curve data) can be found online in \citet{evan_tey_2022_7411579}.

% % Testing having larger size plots and two per page.
% \begin{figure*}[p]
%     \centering
%   \includegraphics[width=.8\textwidth]{figures/example_e.png}\\
%     \includegraphics[width=.8\textwidth]{figures/example_b.png}
% \caption{}
% \label{fig:label_plots}
% \end{figure*}

\begin{figure*}[p]
    \centering
    \includegraphics[width=.8\textwidth]{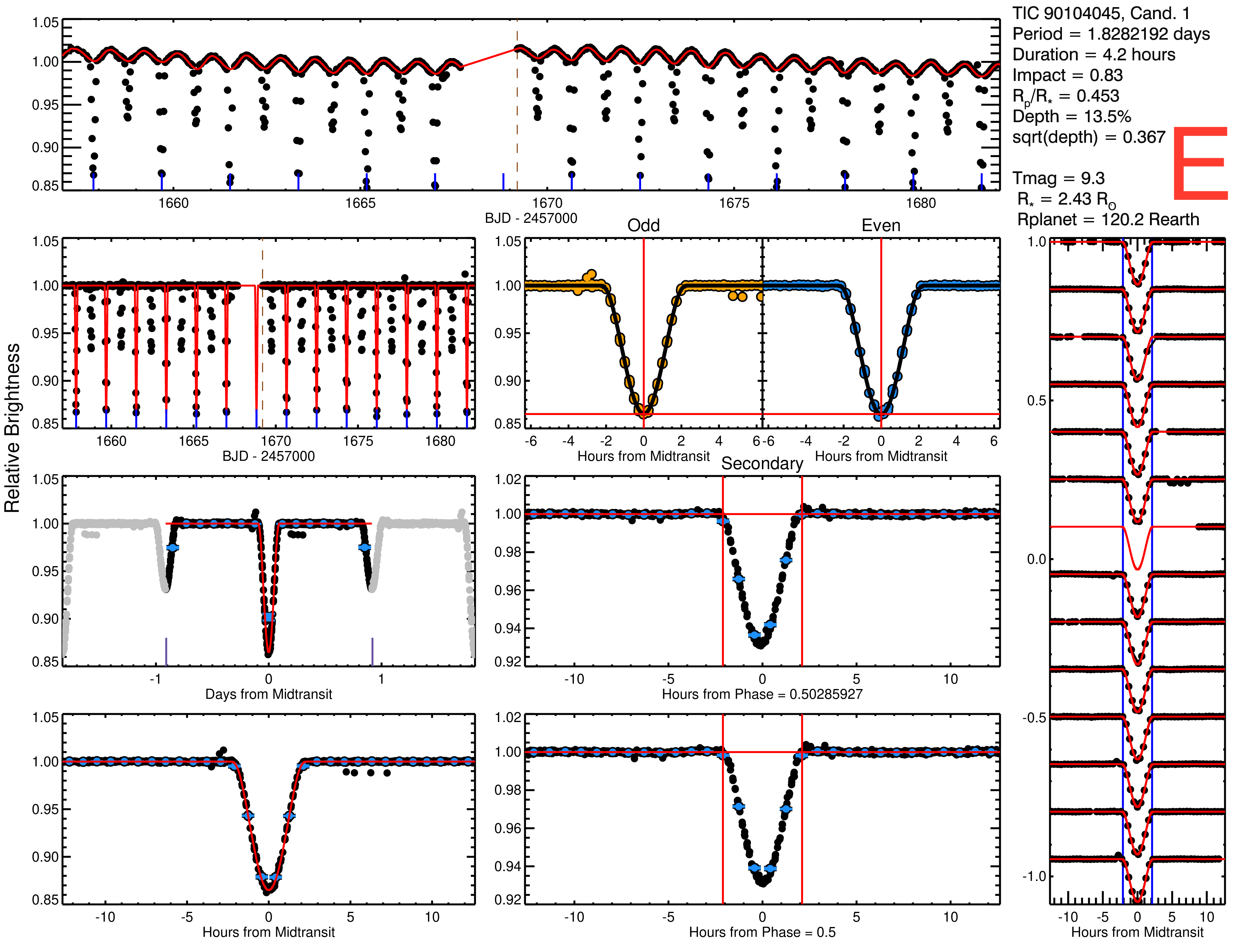}
    \includegraphics[width=.8\textwidth]{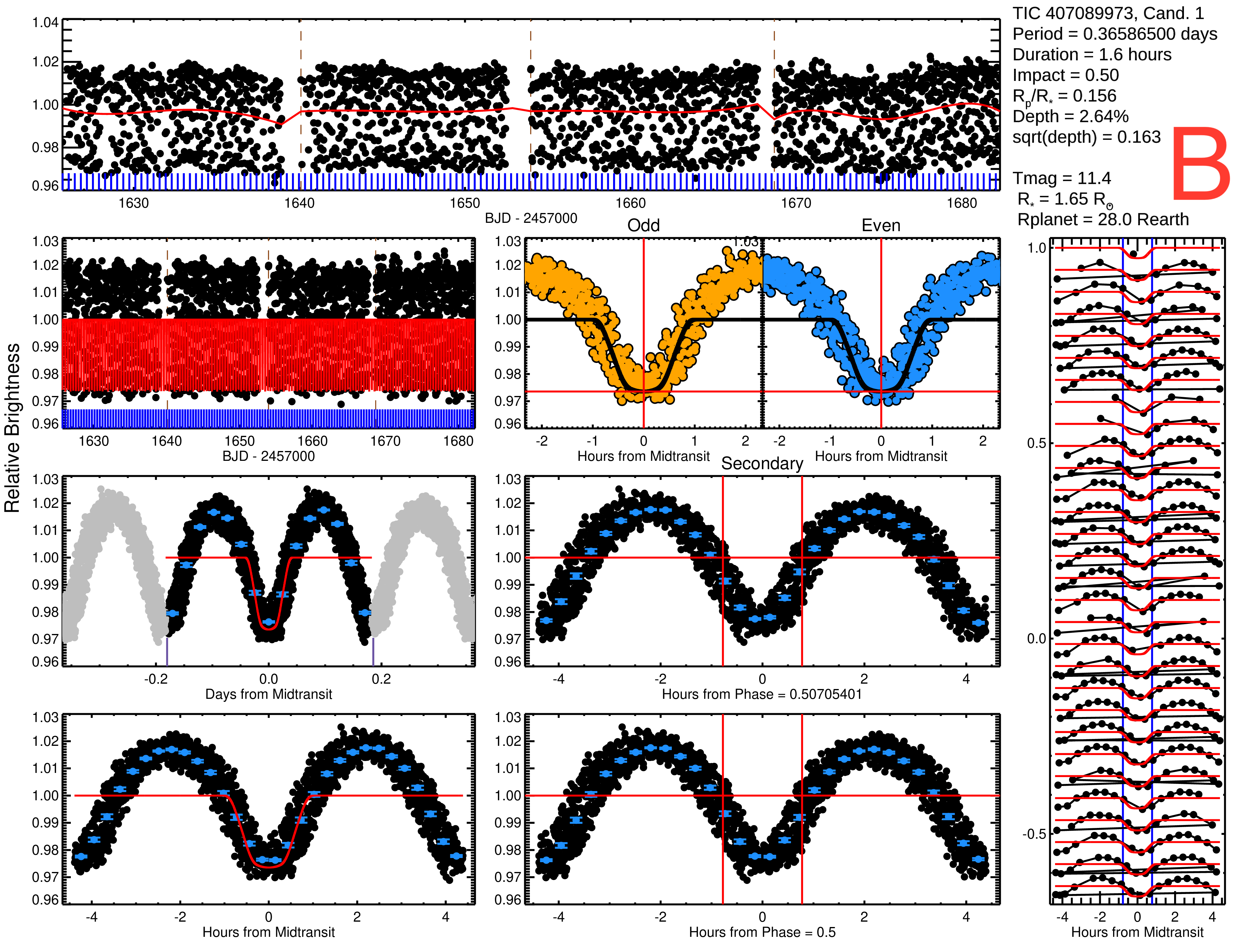}
\end{figure*}
\begin{figure*}[p]
    \centering
    \includegraphics[width=.8\textwidth]{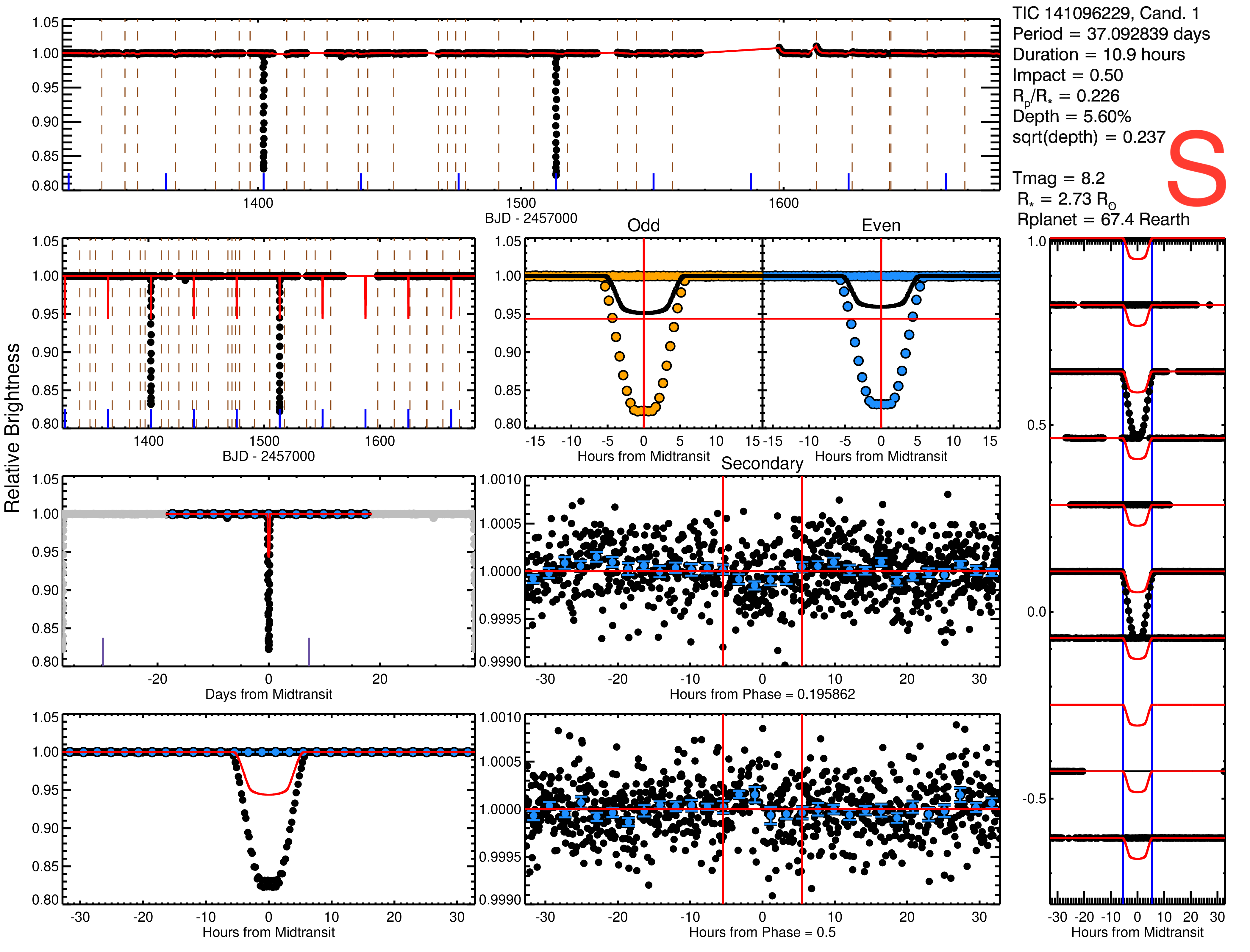}
    \includegraphics[width=.8\textwidth]{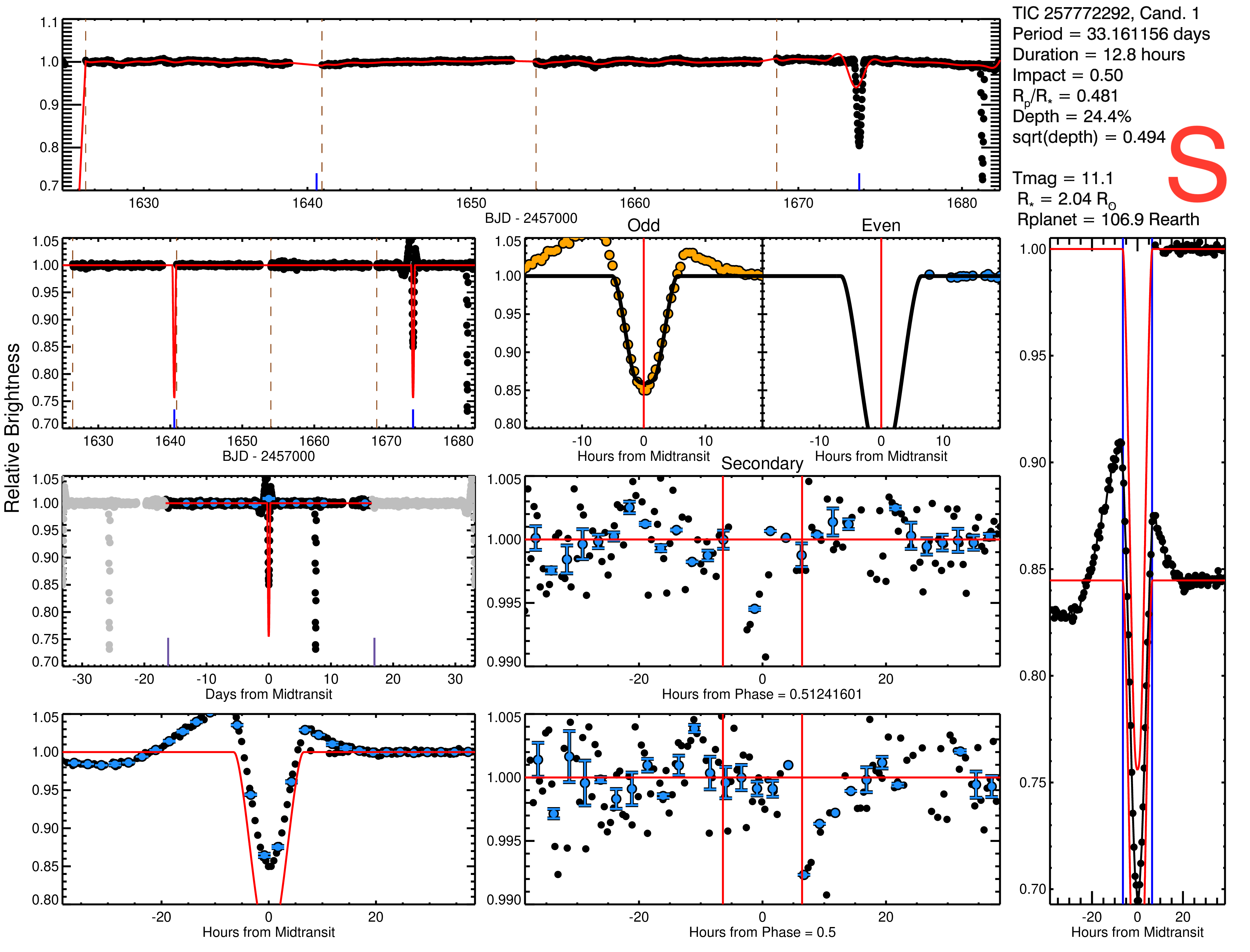}
\end{figure*}

Following common practice in ML, we randomly separate the dataset into a training, validation, and test set. The model is initially fit on the training set, a set of examples used to fit the parameters of the model. Next, the validation set provides a measure of predictive accuracy and model fit. The validation set consists of examples that the model has not seen in the training set and allows for optimization of the architecture and hyperparameters. Lastly, after the model architecture and hyperparameters are finalized, the test set is used as one last objective test of the model accuracy and fit.

\begin{enumerate}

\item Training set ({\bf 19919 targets}): used for model training. \update{(15414 J + 2102 E + 1681 B + 224 S + 498 N)}

\item Validation set ({\bf 2491 targets}): used to calculate precision, recall, detection threshold for binary classification, and model debugging. \update{(1945 J + 261 E + 198 B + 17 S + 70 N)}

\item Test set ({\bf 2516 targets}): hold-out set used for final evaluation; this set was never used for training or debugging, or any other evaluation. \update{(1970 J + 250 E + 200 B + 34 S + 62 N)}

\end{enumerate}

\begin{figure}
    \centering
    \includegraphics{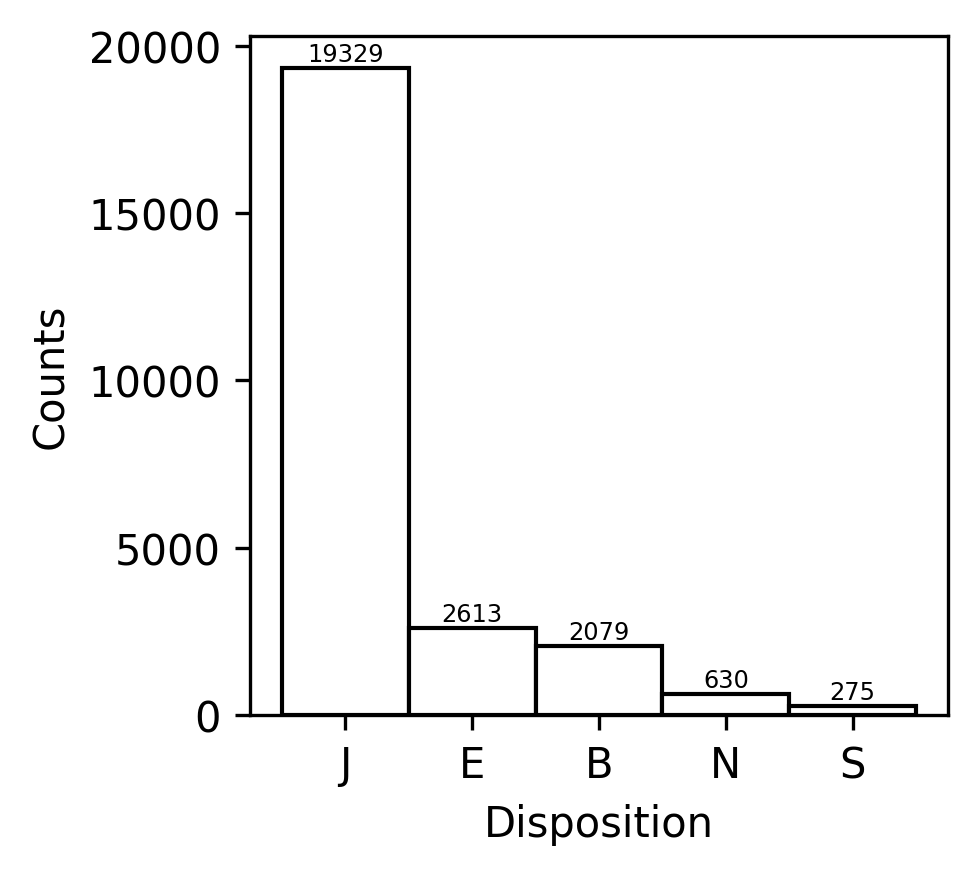}
    \caption{Distribution of labels across our dataset (see Section \ref{labels} for descriptions of each type). As described in Section \ref{label_practice}, some TCEs were assigned fractional B and J labels so these counts have been rounded to the nearest integer.}
    \label{fig:label_counts}
\end{figure}

\begin{figure}
    \centering
    \includegraphics[width=\linewidth]{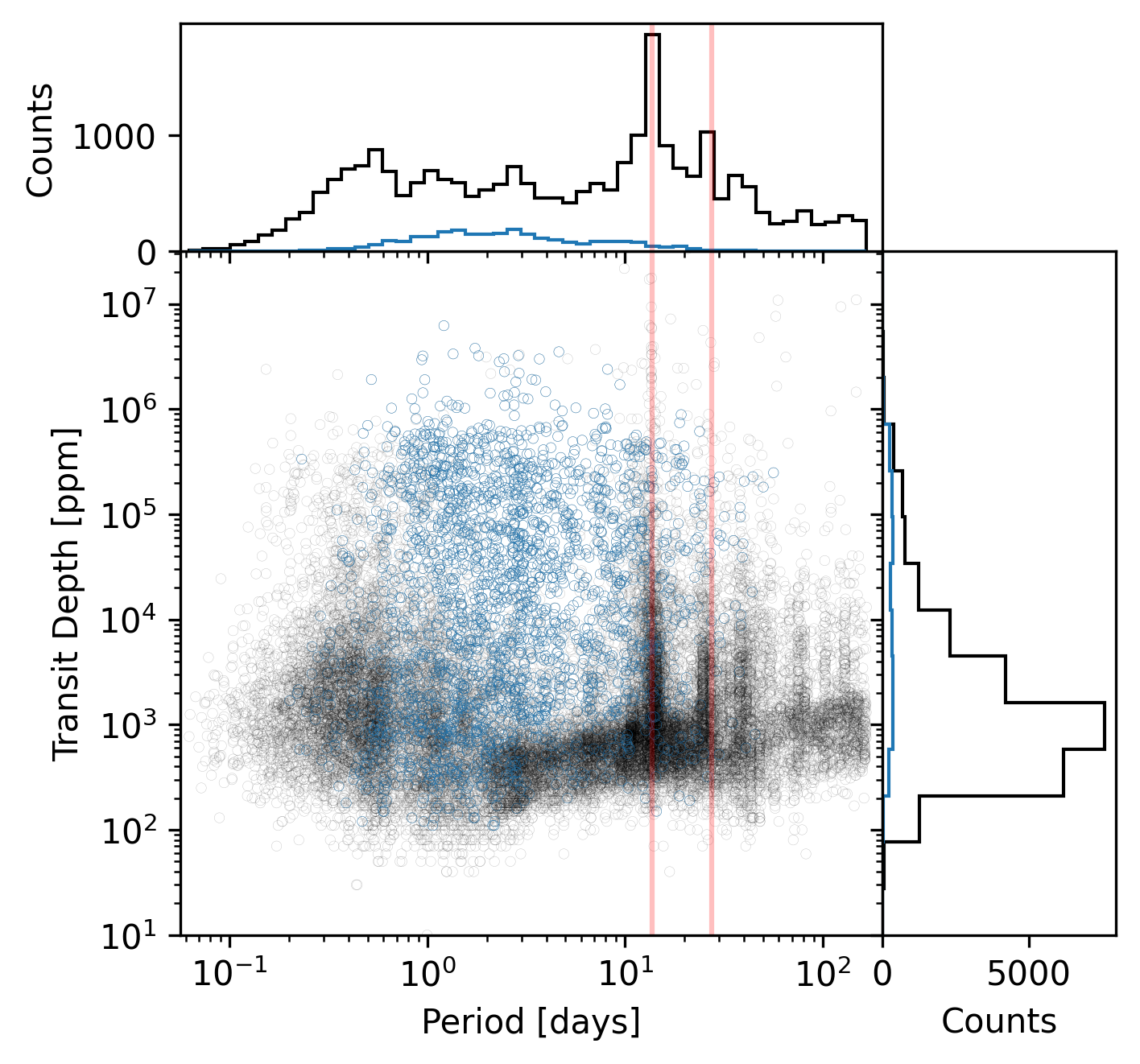}
    \caption{Scatterplot of transit depth vs.~orbital period for our dataset. TCEs with E labels are shown in blue. Red lines mark 13.7 and 27.4, the orbital period and twice the orbital period of TESS.}
    \label{fig:period_v_depth}
\end{figure}

\begin{figure}
    \centering
    \includegraphics[width=\linewidth]{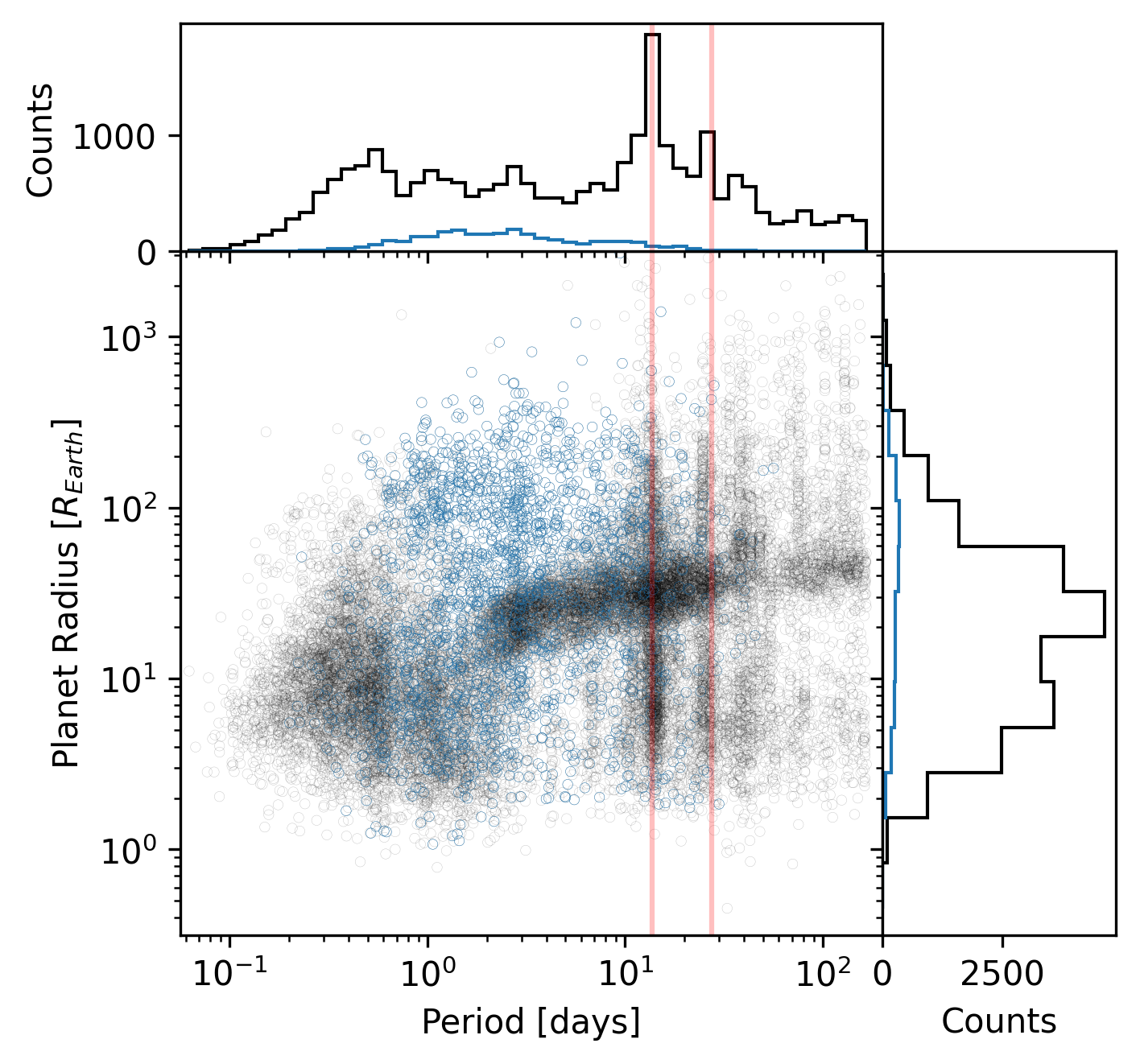}
    \caption{Scatterplot of planet radii vs.~orbital period for our dataset. TCEs with E labels are shown in blue. Red lines mark 13.7 and 27.4, the orbital period and twice the orbital period of TESS.}
    \label{fig:period_v_rad}
\end{figure}

\begin{figure}
    \centering
    \includegraphics[width=\linewidth]{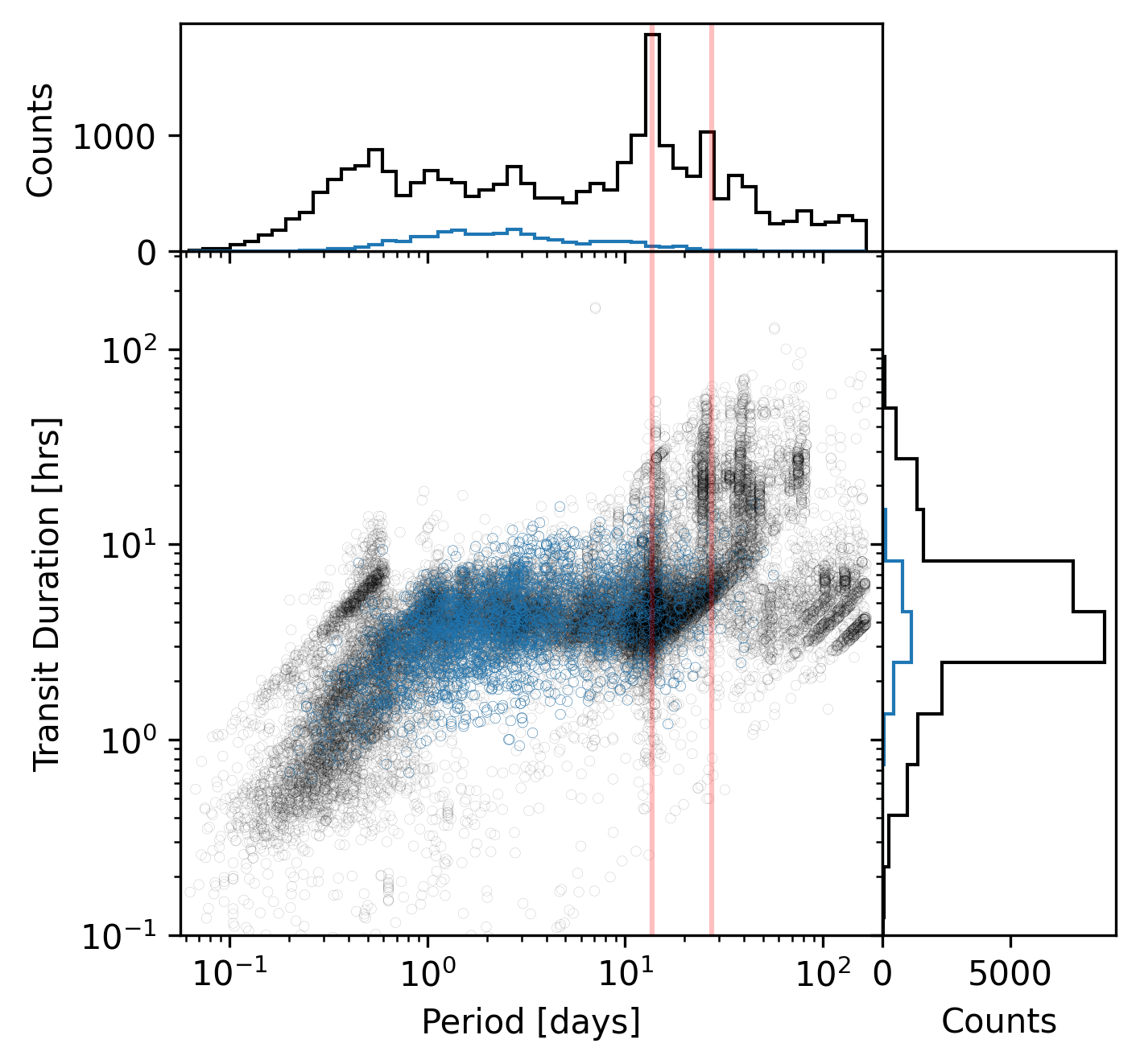}
    \caption{Scatterplot of transit duration vs.~orbital period for our dataset. TCEs with E labels are shown in blue. Red lines mark 13.7 and 27.4, the orbital period and twice the orbital period of TESS.}
    \label{fig:period_v_duration}
\end{figure}

\subsection{Distribution of the labels}

Figure \ref{fig:label_counts} shows the distribution of labels in our training set. Out of the total 24926 labels, the majority are {\bf J} labels (19329). The amount of signals identified as eclipsing objects ({\bf E}, 2613) is comparable to that identified by contact binaries ({\bf B}, 2079). 

We examine the distribution of the fundamental transit parameters (i.e., orbital period, transit depth, estimated planet radius, and transit duration) of the labels in Figure  \ref{fig:period_v_depth}, \ref{fig:period_v_rad}, and \ref{fig:period_v_duration}. Specifically, we compare the parameter spaces resided by the E labels to the other labels. The comparison reveals the following characteristics: (1) a majority number of the TCEs with period smaller than $\sim 0.5$ days are not caused by eclipses; (2) a majority of the shallow events with period longer than 10 days are not caused by eclipses; (3) there is clear pile-up of TCEs at the TESS orbital period and its alias, which are not caused by eclipses; (4) a majority of TCEs with extremely short/long transit duration are not caused by eclipses.

\section{Model input representations}\label{inputrepresentation}

For each TCE, we pass the raw flux time series leading to the detection and all the relevant information describing the detected periodic signal and target star to the neural network.

\subsection{Time series data}

We preprocess the raw flux time series into different input representations before passing them to \texttt{Astronet-Triage-v2}. We use the same basis spline techniques used in QLP, however, the transit signals are masked out based on the BLS-detected period, epoch and duration before the optimal spline is computed. This approach will often prevent over-fitting of the transit signals during the detrending process. To account for different time scales of the stellar variability, we adopt multiple detrending settings to provide \texttt{Astronet-Triage-v2} a more complete view of the light curve noise characteristics. Unlike in QLP, which only uses one set of splines with spacing between 0.3 and 1.5 days to create the final detrended light curves, we use three different settings (0.3, 5.0, and a value which minimizes the Bayesian Information Criterion, \citealt{bic}) to create three different sets of detrended light curves. The light curves detrended with larger spacing are also less likely to over-fit the transit signals with long transit duration.

For each detrended light curve we generate seven different plots or views (see Figure \ref{fig:cnn}). Each view is binned using a robust binning technique to de-weight outliers. During this binning, we also account for the change in exposure time between the Primary and 1st Extended Mission by weighing points according to their exposure time in a given bin. After this, we normalize the binned data so that the minimum value is -1 and the median value is 0. The complete list of views can be found in the source code \footnote{\url{https://github.com/mdanatg/Astronet-Triage/blob/e4ec517b175b2a3dfb74cf6c6e3f5273dd8749c7/astronet/astro_cnn_model/configurations.py##L2254}}. A detailed description of each view type is below:

\begin{itemize}
    \item Global View: The global view uses the full light curve folded on the reported period with 201 bins. In addition to the median values, the view also includes the standard deviations for each bin, a mask indicating whether the bin was empty, and a mask indicating whether the bin falls inside the detected transit.
    \item Local View: The local view uses points within two transit durations of the transit center (for a full timespan of four transit durations), again folded on the reported period. The local view uses 61 bins, and includes standard deviation and mask values like the global view. In addition, we also record the scale factor used in normalization, as a scalar feature.
    \item Secondary View: The secondary view is similar to the local view, but is centered around the most significant secondary transit, determined by performing a grid search\footnote{\url{https://github.com/mdanatg/Astronet-Triage/blob/e4ec517b175b2a3dfb74cf6c6e3f5273dd8749c7/light\_curve\_util/find\_secondary.py##L62}} on the out-of-transit portion of the phase folded view, for a duration equal to the primary transit duration, and selecting the region with the highest signal/noise ratio. This view is accompanied by two scalar features: the normalization scale factor, and the phase of the secondary transit's center.
    \item Local Half-Period View: Similar to the local view, but folded at half the detected period. This view only contains the standard deviation value, since the median value can appear very noisy when folding a transit over a non-transit.
    \item Global Double Period View: Similar to the global view, but folded at twice the period of the global view.
    \item Sample Global Segments: This view contains the entire period (similar to the global view), but showing up to 7 of the folds that contain the most points (ties are broken at random). Each fold is accompanied by a mask indicating whether the bin contains any points. If the light curve contains fewer transits, the extra views remain empty. Each fold is independently binned with 201 bins.
    \item Sample Local Segments: Similar to the sample global segments, this view contains the transit center of up to 4 of the folds that contain the most points (ties are broken at random), for a total of 8 folds. Each fold is independently binned with 61 bins.
\end{itemize}

\subsection{Scalar data}\label{sec:scalar_data}
We also use scalar values that describe characteristics of the transit, host star and the light curve itself. Transit features include period in days ($P$), transit duration in days  ($T_{\rm dur}$), transit depth ($\delta$), and the number of full periods observed in the flux-time series ($n_{\rm folds}$), while host star features include TESS magnitude ($T_{\rm mag}$), mass in $M_\odot$, and radius in $R_\odot$. The host star features are directly extracted from the TESS Input Catalog v8.2 \citep{Paegert2021}. 

For TCEs without stellar radii in the catalog, we perform a rough estimate using a Bayesian estimate of the distance \citep{bailerjones}, apparent magnitude (either Gaia G, Bp, and Rp, or Gaia G and 2MASS K if Bp and Rp are unavailable), and color/temperature and color/bolometric corrections from MIST models \citep{choi}. In brief, we estimate the temperature and bolometric correction from either the target's Bp-Rp or G-K colors, use the bolometric correction to estimate the target's apparent bolometric magnitude, use the estimated distance to the target to convert to an absolute magnitude, convert to bolometric luminosity, and solve for the stellar radius from the temperature and luminosity via the Stefan Boltzmann Law. In our testing, we were able to determine radii within about 10\% of the TIC values when present, and provided radius estimates for $\sim$ 2400 from the $\sim$ 2800 TCEs missing stellar radii in our dataset.      

Light curve features include the total number of points. Each scalar value is normalized to be zero mean and unit variance across the dataset, except for $n_{\rm folds}$ which is truncated to a maximum value of 100 and a log-scaled to fit between 0 and 1. In addition, we also include as scalar inputs the detected phase of the secondary eclipse, as well as the calculated scaling factor when normalizing the global, local and secondary views.

\section{Neural network architecture}\label{architecture}

\begin{figure*}
    \centering
    \includegraphics[width=\linewidth]{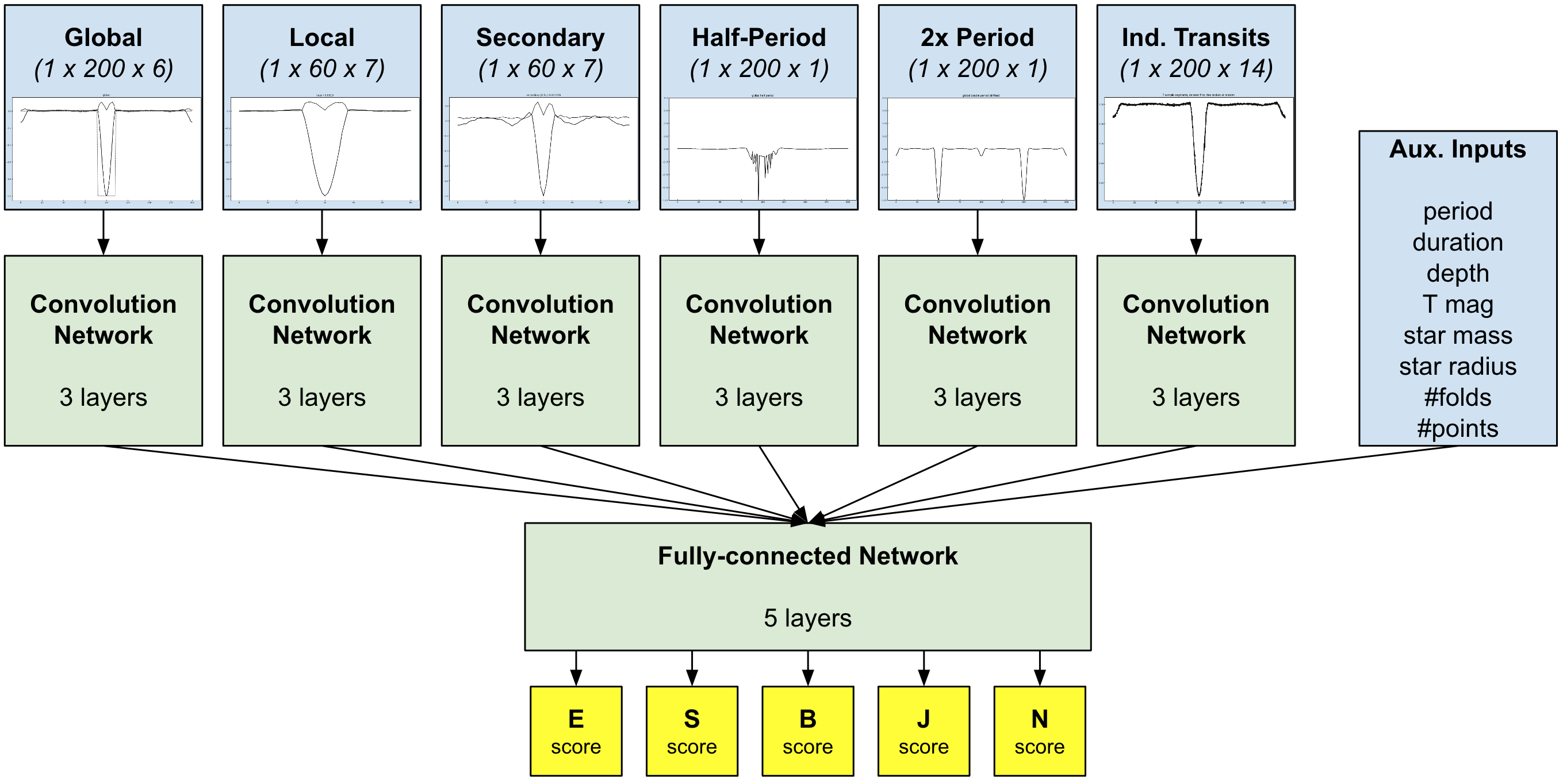}
    \caption{\texttt{Astronet-Triage-v2} neural network architecture. 
    \label{fig:cnn}}
\end{figure*}

Our model uses a convolutional neural network architecture derived from \texttt{Astronet}. The high level architecture is shown in Figure \ref{fig:cnn}.

\begin{figure}
    \centering
    \includegraphics[width=\linewidth]{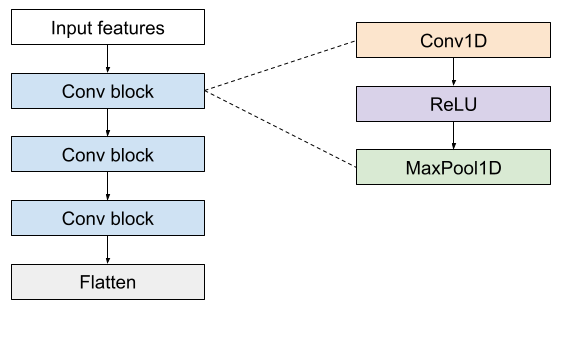}
    \caption{Structure of a CNN tower. Each convolution tower has 1 to 4 blocks. Each block has 1 to 4 layers. 
    \label{fig:cnn-tower}}
\end{figure}

Each time series feature is grouped together with similar features and then passed through a separate convolutional tower. For example, the global view flux is grouped together with the standard deviation of the global view, so that they form a 2-channel, 1-dimensional image. The structure of a convolutional tower is shown in Figure \ref{fig:cnn-tower}. Each tower consists of convolutional layers with Rectified Linear Unit (ReLU) activation, alternating with pooling layers. The pooling layers aggregate neighboring pixels, in effect increasing the field of view of the subsequent convolutional layer.

\begin{figure}
    \centering
    \includegraphics[width=\linewidth]{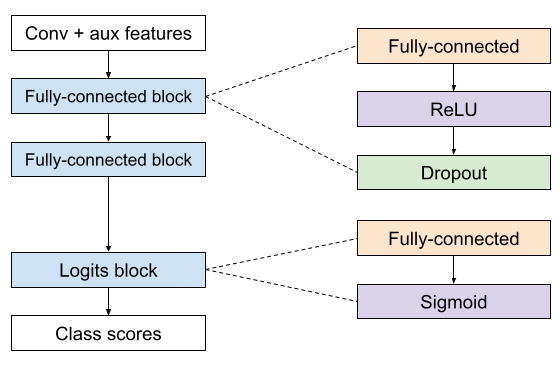}
    \caption{Structure of the fully-connected tower. 
    \label{fig:fc}}
\end{figure}

The output of each convolutional tower is flattened into a vector shape. The flattened outputs from all towers are concatenated together with the auxiliary inputs to form the input into the next section of the network, the fully-connected tower, whose structure is shown in Figure \ref{fig:fc}. The fully-connected tower is composed of several fully-connected neural network layers, alternating with dropout layers. The dropout layers randomly set inputs to zero, and serve a role of regularization, to mitigate over-fitting. The dropout layers are only active during training. The final layer has five outputs, and uses a sigmoid activation function, so that its output is in the interval \([0..1]\). Each of the five outputs corresponds to one of the five labels.

The various hyper-parameters of each network can be found in the configuration file included with the source code.\footnote{\url{https://github.com/mdanatg/Astronet-Triage/blob/e4ec517b175b2a3dfb74cf6c6e3f5273dd8749c7/astronet/astro\_cnn\_model/configurations.py}} The hyper-parameters are tuned using Vizier \citep{google_vizier, oss_vizier} by minimizing the loss on the validation set.

\subsection{Training}\label{training}

We train the model using the Adam, a popular variant of stochastic gradient descent optimization \citep{adam}, for 20,000 steps. The complete set of training parameters can be found in the code \footnote{\url{https://github.com/mdanatg/Astronet-Triage/blob/e4ec517b175b2a3dfb74cf6c6e3f5273dd8749c7/astronet/astro\_cnn\_model/configurations.py##L2254}}.

For the loss function we use binary cross-entropy loss\footnote{See \url{https://www.tensorflow.org/api_docs/python/tf/keras/losses/BinaryCrossentropy} for the implementation and \citet{10.2307/2984087} and \citet{2018AJ....155...94S} for more information}. Notably, this means that the model is not trained to choose between the five labels exclusively. Instead, it produces independent scores for each label, so a model is free to assign high scores for both ``E'' and ``J'' labels, for instance. This loss function enables us to assign weighted labels to uncertain examples (e.g. 50 percent ``B'', 50 percent ``J''). The weight is determined as follows: if a target had a single label (as resulting from the group resolution, or if the vote was unanimous), the weight is 1.0; if the target had multiple votes, the weight is the maximum number of votes for any label divided by the total number of votes. This means targets for which a label didn't receive a majority of votes are weighted less.

We don't apply data augmentation, although that is something we intend to do in future work (see Section \ref{sec:future_improvements}).

\subsection{Prediction and ensembling}

As a multi-class classifier, our model outputs a prediction score for each label. Predictions where the ``E'' label score exceeds a threshold chosen beforehand are considered to predict the label ``E''. Otherwise, the model is considered to predict the label with the highest prediction score.

We then construct an ensemble of 10 models trained separately (hence with different initial weight values, and different shuffling of the input data). The compound prediction of the ensemble is constructed as follows:

\begin{enumerate}

\item If any of the models predicts ``E'', then the ensemble prediction is also ``E''.

\item Otherwise, the ensemble prediction is the label predicted by a majority of models, with ties broken at random.

\end{enumerate}

Although the model predicts five different labels, we are primarily interested in the ``E'' label. The other labels are mainly used at training, to encourage the network to learn natural representations. We found that the extra labels greatly help understand a model's predictions, as well as validate whether the model does indeed create correct internal representations.

\section{Results}\label{results}

Here we report the results of our ML activity predictions. First we discuss the metrics we used to evaluate the performance and then we summarize how the different models performed on each dataset.

\update{The two primary metrics we use to evaluate our performance are precision and recall. The precision, or reliability, of a model on a labelled dataset is the number of true positives divided by the number of true positives and false positives. Recall, or completeness, is the number of true positives divided by the number of true positives and false negatives. As we are interested in ``E'' labels as potential planet candidates, they generally are used as the ``positive'' class. In this context, a high precision means our model outputs fewer false positives, meanwhile a high recall means successful recovery of more planet candidates (fewer potential planets lost by \texttt{Astronet-Triage-v2}). Since labels are determined by comparing output prediction scores against a chosen threshold, each specific threshold yields its own precision and recall. When plotted over many different thresholds, we can form a precision-recall curve (see Figure \ref{fig:s33_pr}). By taking the area under the precision-recall curve (AUC-PR), also known as the average precision, we can characterize our model's overall performance and compare against other models with the highest achievable value being a 1.}

\subsection{Performance on validation and test sets} \label{sec:validation_test}

On the validation dataset we obtained an AUC-PR value of ~0.977. The model achieves 100\% recall at 41\% precision, at a prediction threshold of 0.0105. If we increase the threshold to 0.215, we obtain 96.9\% recall at 79.8\% precision.

On the test set, we obtained an AUC-PR value of ~0.965. The model achieves 100\% recall at 15\% precision, at a prediction threshold of 0.0005. This suggests the test set contains more difficult examples (possibly incorrect ones). With the thresholds suggested by the validation set, we obtain 99.6\% recall at 39.7\% precision for the 0.0105 threshold, and respectively 97.2\% recall at 75.7\% precision for the 0.215 threshold.

\subsection{Generalizing to TESS 1st Extended Mission data}\label{sec:em_performance}

We explore the adaptability of our network, and the generalization of training on non-uniform datasets in this section. In practice, models like \texttt{Astronet-Triage-v2} are trained on previously observed sectors with a goal of classifying new observations taken by TESS in the future. Since noise characteristics and TESS observation strategy can change sector-to-sector, it is important that our models generalize well to new data.

Nearly 90\% of our total training dataset comes from the TESS Primary Mission, so we use QLP data from TESS 1st Extended Mission (Sector 33, observed during Year 3 from UT 2020 December 17 -- UT 2021 January 13) to test how our model generalizes to unseen or out-of-distribution data.

Following the QLP convention, we ran a BLS search and \texttt{Astronet-Triage-v2} on the full multi-sector light curves (including both Primary Mission and 1st Extended Mission data) for each star. Of the discovered TCEs, we selected a random sample of 759 targets with $T_{\rm mag} < 11$ from camera 1 and 590 targets with $11 < T_{\rm mag} < 13.5$ from camera 2. Due to the TESS pointing strategy, we focus on these cameras because their light curves have roughly equal amounts of Primary vs.~1st Extended Mission observations. The magnitude ranges also allow us to compare performance on stars in different brightness bins.

One of our vetters (CH) independently labeled all 1349 TCEs before evaluation, among which, 255 TCEs were assigned an E label.

To better understand our ability to generalize, we apply the following models to the Sector 33 dataset: \texttt{Astronet-Triage}, the fully trained \texttt{Astronet-Triage-v2}, and three independent instances of the \texttt{Astronet-Triage-v2} architecture trained on different subsets of our original TCE dataset (Section \ref{observation_descrip}). 

These three separate training sets were formed by splitting our original training set on observation year, meaning roughly 40\% went into training the Y1 model, 50\% into the Y2 model, and 10\% into the Y3 model. The differences between these datasets are described in Section \ref{tess_data}, but briefly: Both the Y1 and Y2 datasets feature brighter stars, but the Y1 dataset were only taken from Sector 13, so they cover a small region of the Southern ecliptic hemisphere. The Y2 dataset, on the other hand, were selected more uniformly and cover most of the Northern ecliptic hemisphere. Neither has much overlap in sky coverage with the evaluation set (the 1349 Sector 33 TCEs) -- Y1 having little overlap and Y2 having none. Both datasets also have much shorter observation baselines than the evaluation set, and finally, due to the change in TESS momentum dump strategy, the Y1 dataset also differs from the evaluation set in noise characteristics. The Y3 dataset bears the most similarity to the evaluation set in terms of data characteristic. It is, however, much smaller than the other datasets. Altogether, these different datasets and models provide useful views at our ability to generalize to data that can be fairly different from the training data.

Since \texttt{Astronet-Triage} only distinguishes between transit-like and non-transit-like, it's trained to give high scores TCEs we consider E- or S- labeled. As \texttt{Astronet-Triage-v2} provides independent E and S scores, we choose remove all S-labeled data from precision and recall calculations for a simple direct performance comparison with \texttt{Astronet-Triage}. This leaves us with 1315 TCEs.

Precision and recall numbers split across \texttt{Astronet-Triage} and \texttt{Astronet-Triage-v2} for each camera can be seen in Table \ref{tab:s33_results}. In both cameras we see that for similar (or better) levels of precision, \texttt{Astronet-Triage-v2} provides better recall than \texttt{Astronet-Triage}, \update{with a slightly more pronounced effect in camera 2 (fainter targets). In other words, for the same amount of human vetting time, \texttt{Astronet-Triage-v2} would recover more potential planets than \texttt{Astronet-Triage}}. 

\begin{table}
\caption{Performance on previously unseen S33 data \label{tab:s33_results}}
\begin{tabular}{l|c|c|c|c}
    Model & Cam & Threshold & Precision & Recall \\
    \hline
    \texttt{Astronet-Triage-v2} & 1 & 0.0105 & 0.64 & 0.98 \\
    \texttt{Astronet-Triage-v2} & 2 & 0.0105 & 0.53 & 1.00 \\
    \texttt{Astronet-Triage-v2} & 1 & 0.215 & 0.89 & 0.91 \\
    \texttt{Astronet-Triage-v2} & 2 & 0.215 & 0.84 & 0.99 \\
    \texttt{Astronet-Triage} & 1 & 0.08 & 0.89 & 0.85 \\
    \texttt{Astronet-Triage} & 2 & 0.08 & 0.82 & 0.90 \\
\end{tabular}
\end{table}

\begin{figure}
    \centering
    \includegraphics[width=0.5\textwidth]{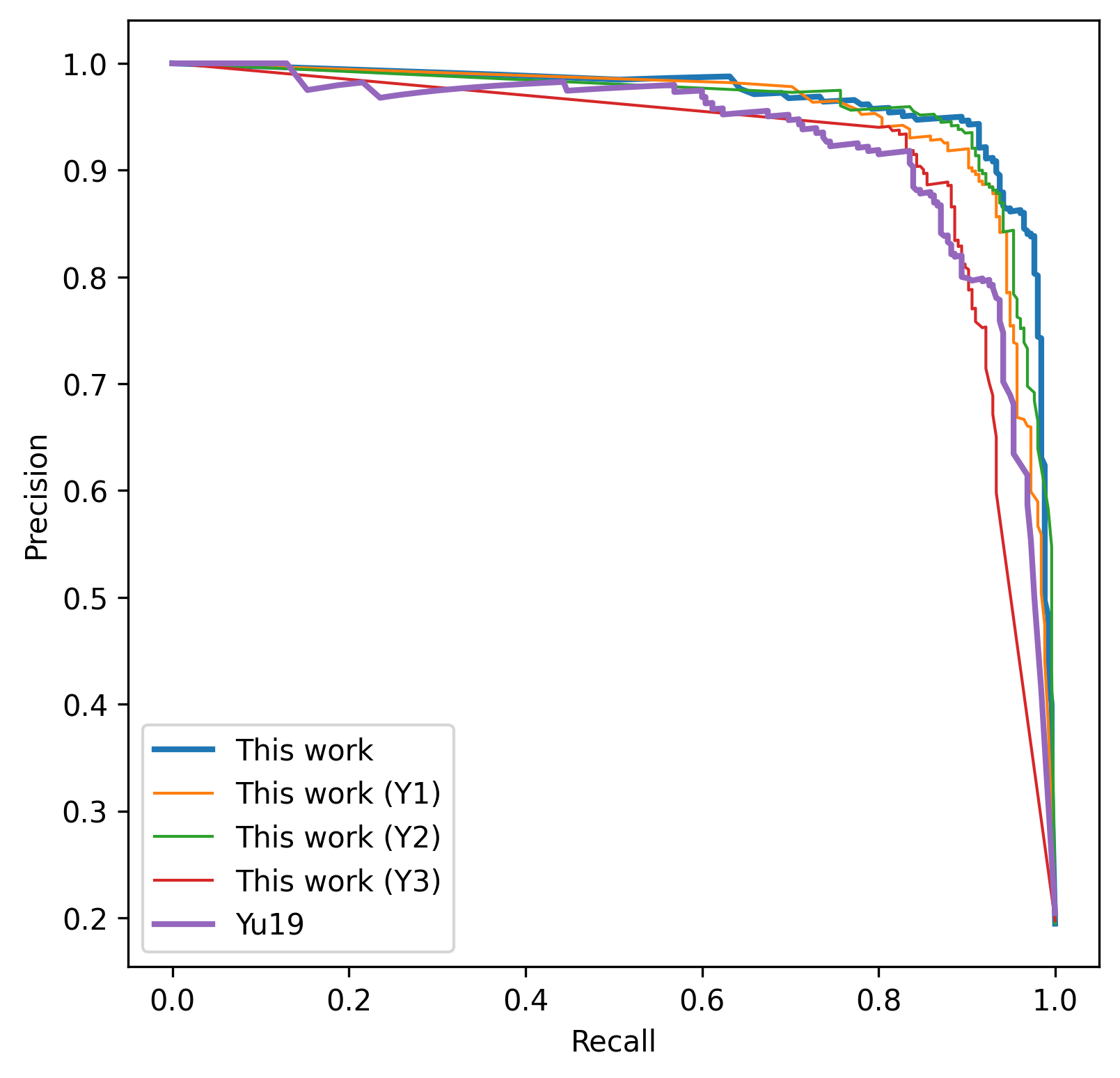}
    \caption{Precision vs.~recall for 1315 TCEs selected from Sector 33 of the 1st Extended Mission. 
    Since \texttt{Astronet-Triage} \citep{2019AJ....158...25Y} only distinguishes between transit-like and non-transit-like, it gives high scores to TCEs we either consider to have E or S labels. For a more direct comparison to \texttt{Astronet-Triage-v2}, we choose to ignore all S-labeled TCEs when calculating precision and recall.
    We see that across all levels of recall, \texttt{Astronet-Triage-v2} provides higher precision even when trained only on Primary Mission data taken during Y1 or Y2. Although the Y3 dataset bears the most resemblance to the S33 evaluation set here, the size of the Y3 dataset is only $\sim 2500$, so the Y3-trained model doesn't quite reach the performance of the other models.
    }
    \label{fig:s33_pr}
\end{figure}

The full precision-recall curves across all TCEs (ignoring S-labeled TCEs) are shown in Figure \ref{fig:s33_pr}. Across the board we see that \texttt{Astronet-Triage-v2} (trained on the full training set) improves on \texttt{Astronet-Triage} with AUC-PR scores of 0.961 and 0.927. We also see that the models trained only on Y1, Y2, and Y3 data perform similarly to \texttt{Astronet-Triage} with AUC-PR scores of 0.954, 0.960, and 0.917 respectively. Even though the Y1 and Y2 versions of the models don't use any 1st Extended Mission training data, we see they're still able to perform highly in S33 (which occurred during Y3). This supports \texttt{Astronet-Triage-v2}'s ability to generalize to future sectors.

\subsection{Performance on the TOI catalog}\label{sec:toi_performance}

The TESS Objects of Interest (TOI) catalog \citep{guerrero}, which lists the planetary candidates detected by \TESS, is a useful benchmark for high-confidence E or S labels. A good model should label all TOI entries as E or S, since humans have inspected each entry and considered them to be high-probability planetary candidates (allowing for single-transit events).

On 2022 April 21 we downloaded the TOI catalog with light curve data through Sector 47. We also use information from TESS Follow-up Observing Program (TFOP) Sub Groups 1 and 2 (SG1 \& SG2), \update{which use ground-based photometry and reconnaissance spectroscopy to follow-up on TOIs and help filter out false positives}. After keeping only planet candidates (PCs; meaning TOIs that were not ruled out as false positives with follow-up observations) and validated / confirmed / known planets (Ps), we have a dataset of 4140 targets.

\begin{figure}
    \centering
    \includegraphics[width=0.5\textwidth]{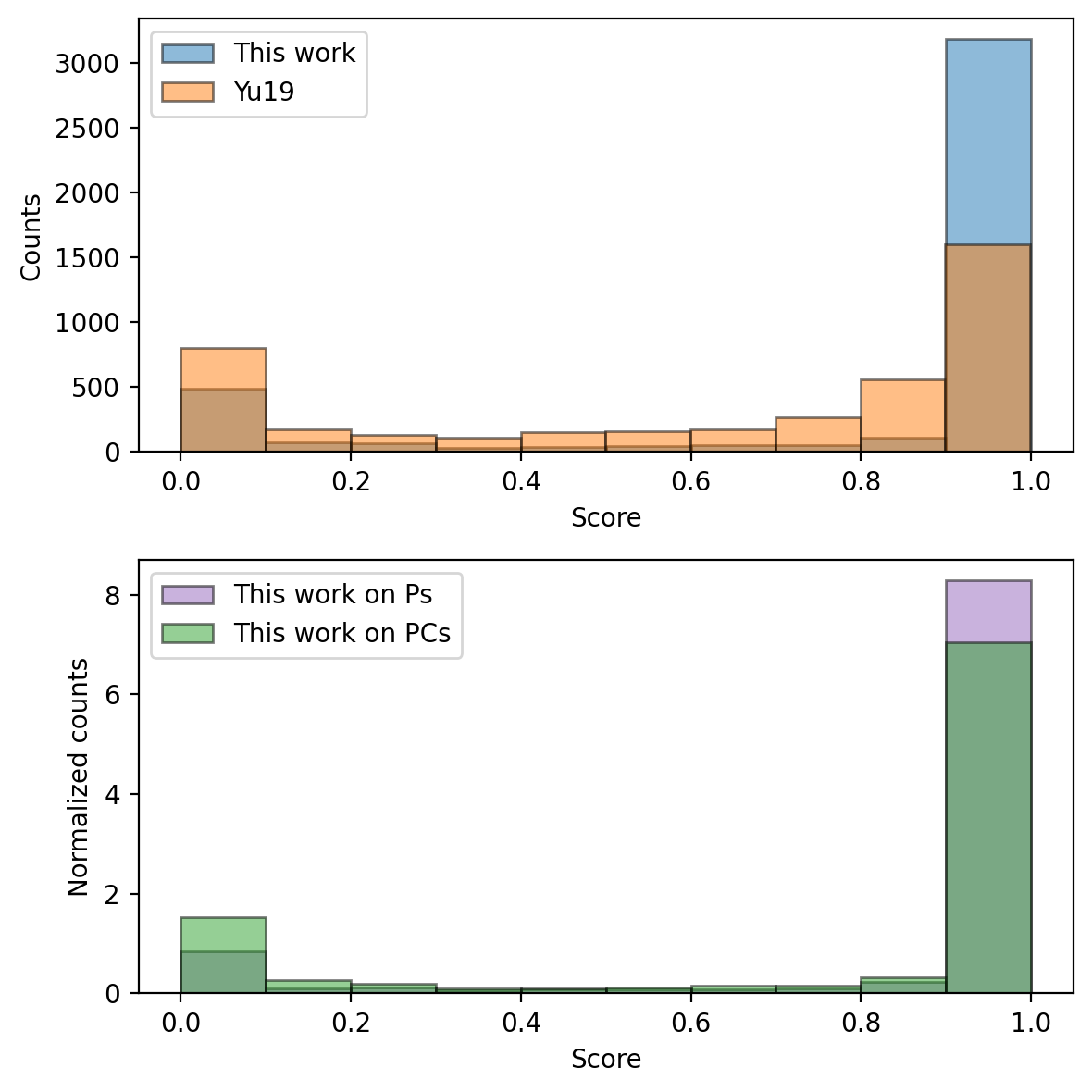}
    \caption{Top: Distribution of E score between this work and \texttt{Astronet-Triage} \citep{2019AJ....158...25Y} on the whole TOI dataset. Bottom: Distribution of E scores from this work when the dataset is separated into Planets (P, validated, confirmed, and known planets) and Planet Candidates (PC, TOIs that are not validated, confirmed, or known planets, and were also not identified as false positives with follow-up observations).}
    \label{fig:toi_hist}
\end{figure}

\begin{figure}
    \centering
    \includegraphics[width=0.5\textwidth]{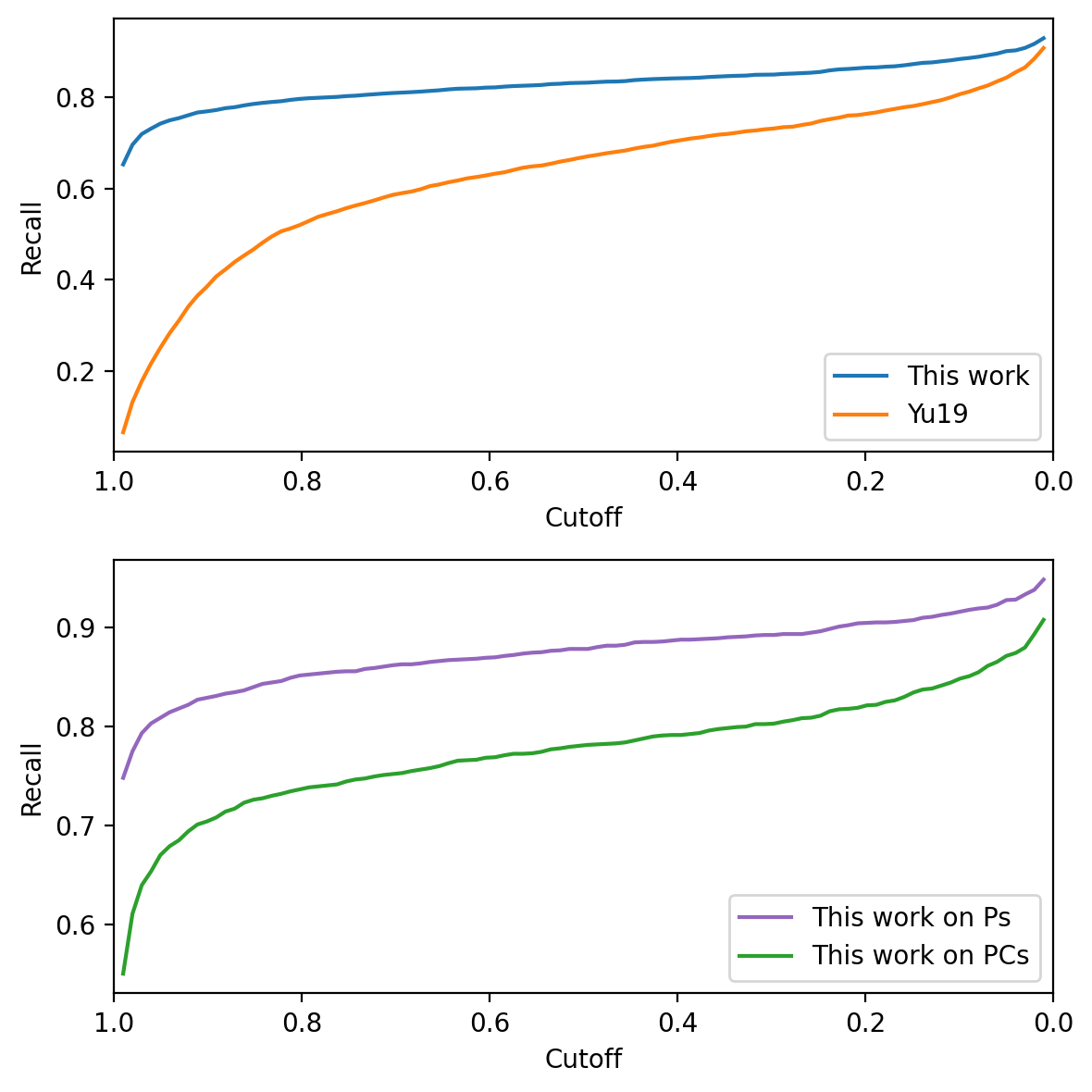}
    \caption{Top: Recall as a function of cutoff threshold between this work and \texttt{Astronet-Triage} \citep{2019AJ....158...25Y}. For \texttt{Astronet-Triage-v2} we choose to focus on just E scores even though some TOIs are true S labels. Bottom: \texttt{Astronet-Triage-v2} recall as a function of cutoff threshold when the dataset is separated into Planets (P, validated, confirmed, and known planets) and Planet Candidates (PC, TOIs that are not validated, confirmed, or known planets, and were also not identified as false positives with follow-up observations).}
    \label{fig:toi_recall}
\end{figure}

After evaluating all TOI signals with \texttt{Astronet-Triage-v2}, Figure \ref{fig:toi_hist} shows the distribution of E scores. Figure \ref{fig:toi_recall} shows the recall rate at different cutoff threshold levels. We see that 93\% of the TOIs have E scores $>0.0105$ and as we increase the cutoff to $0.215$, \texttt{Astronet-Triage-v2} passes 86\% of the TOIs. We also see improved \texttt{Astronet-Triage-v2} performance on known, confirmed, or validated planets (Ps) compared to the planet candidates (PCs) across the board.

For comparison, we also ran \texttt{Astronet-Triage} on all TOI signals. Using a threshold of 0.09, as was originally used in QLP, \texttt{Astronet-Triage} recovers 3349 TOIs. Using the dataset from Section \ref{sec:em_performance}, we find a precision-matching threshold of 0.2 for \texttt{Astronet-Triage-v2}. \update{By finding the threshold of equal precision, we can compare TOI recovery at a constant rate of human vetter work. At this threshold, 3577 TOIs are recovered. In other words, at least 200 TOIs are saved by using \texttt{Astronet-Triage-v2} in place of \texttt{Astronet-Triage} without introducing more false positives to human vetters.}

Some important caveats to note:
\begin{itemize}
    \item The TOI catalog does include single-transit events. \texttt{Astronet-Triage-v2} is trained to give these S rather than E labels. Rather than keeping separate cutoffs for S and E scores, for simplicity we choose to focus on E scores in reported recalls. This gives it a slight disadvantage in terms of recovery numbers, though we leave them in the dataset for fairer comparison to \texttt{Astronet-Triage} which gives a score for transit-like (periodic or single-transit) versus not transit-like.
    \item TOIs can also come from the SPOC pipeline, which processes 2-minute cadence light curves. For both \texttt{Astronet-Triage-v2} and \texttt{Astronet-Triage}, QLP light curves are binned down to 30 or 10 minutes, so some signals may not be detectable (e.g. due to low signal-to-noise in the binned light curve) and should be assigned J labels. This contributes partially to the lower recall numbers seen at the cutoffs from Section \ref{sec:validation_test}.
    \item \update{Only 130 TOI host stars appear in our dataset of $\sim$25,000, 100 of which were in the training set. We also conducted this analysis with those TOIs removed and saw similar results.}
\end{itemize}

% \begin{enumerate}
%     \item produced catalog from SG1 TOI list with sg1 and sg2 info. 
%     \item explain sg1 (get NEBs) and sg2 (get on target binaries)
%     \item only kept rows with one of these dispositions: "CPC, CPC-, CPC?, KP, P, PC, PPC, VPC, VPC-, VPC-?, VPC?, VPC+, VPC+?" (candidates, validated/confirmed/known planets)
%     \item for everything below I think we want to break down results by PC or not PC?
%     \item plot: recall as a function of cutoff
%     \item cite a couple specific points on this plot
%     \item plot: histogram of E scores
%     \item discussion about worst performers:
%     \item someone should probably investigate more into why they failed (Evan can make a sheet somewhere if anyone wants to look)
%     \item another caveat here is that we're underreporting our recall score because there are things like duplicate rows in the SG1 sheet (eg when period isnt known)? (single transits and period aliasing are just sads).
% \end{enumerate}
% ET: old text here. I think chelsea did this comparison? These look/sound a lot better than the current results I'm getting? I see something like 90% recall
% We obtained 99.0\% recall (19 false negatives) at a low threshold of 0.03. At a higher threshold of 0.31, we obtained 98.5\% recall (30 false negatives). Interestingly, after revisiting 6 of the entries labeled negative by the model, we have confirmed that they are indeed true negatives and removed them from the catalog.

\section{Discussion}\label{discussion}

% Michelle: can talk about importance for demographics and occurrence rates, i.e. automated vetting that can be well characterized, with measured completeness
\subsection{Use in producing the TOI catalog}
\update{A large piece of motivation for this work has been improving on \texttt{Astronet-Triage} so fewer planet candidates are lost when searching for TOIs via QLP. After signal detection via BLS, Astronet is one of the finals triage steps before candidates are passed along to human TOI vetters and potentially promoted to TOIs \citep{guerrero}. Based on the results in Section \ref{results} we expect \texttt{Astronet-Triage-v2} to save many planet candidates that would otherwise be lost without adding false positives and increasing the hours needed for human TOI vetting. Starting in Sector 34, early versions of \texttt{Astronet-Triage-v2} officially replaced \texttt{Astronet-Triage} within QLP. While \texttt{Astronet-Triage-v2} takes step towards a more automated process, it is still not developed enough for population statistics (for a deeper discussion see Section \ref{sec:population_statistics}).}

\subsection{What is limiting our precision?}

\begin{figure*}
    \centering
    \includegraphics[width=0.8\textwidth]{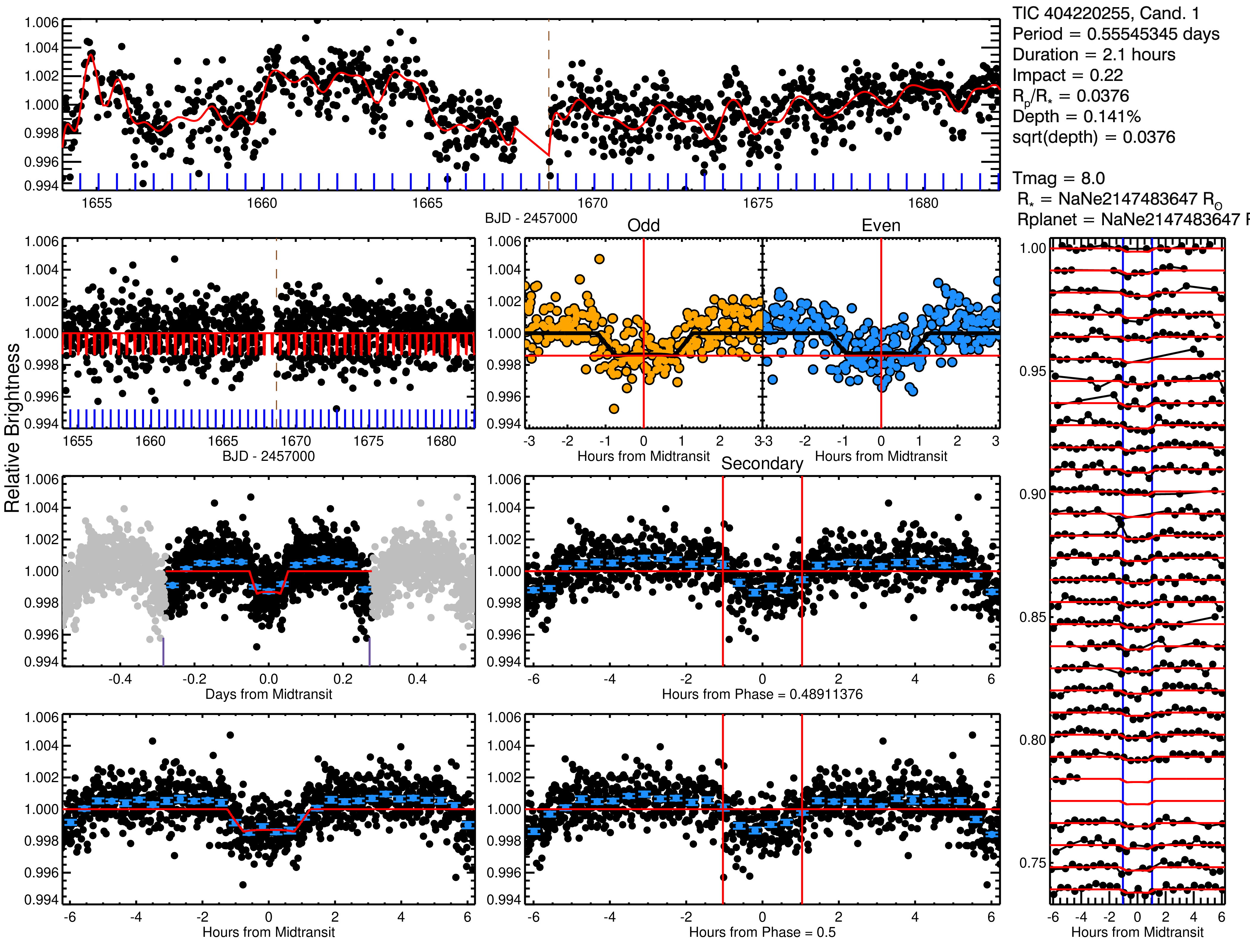}
    \caption{Example of borderline pattern. The true label for this example is ``E'', but the folded light curve appears very similar to a ``B''.}
    \label{fig:hardexampleblike}
\end{figure*}

\begin{figure*}
    \centering
    \includegraphics[width=0.8\textwidth]{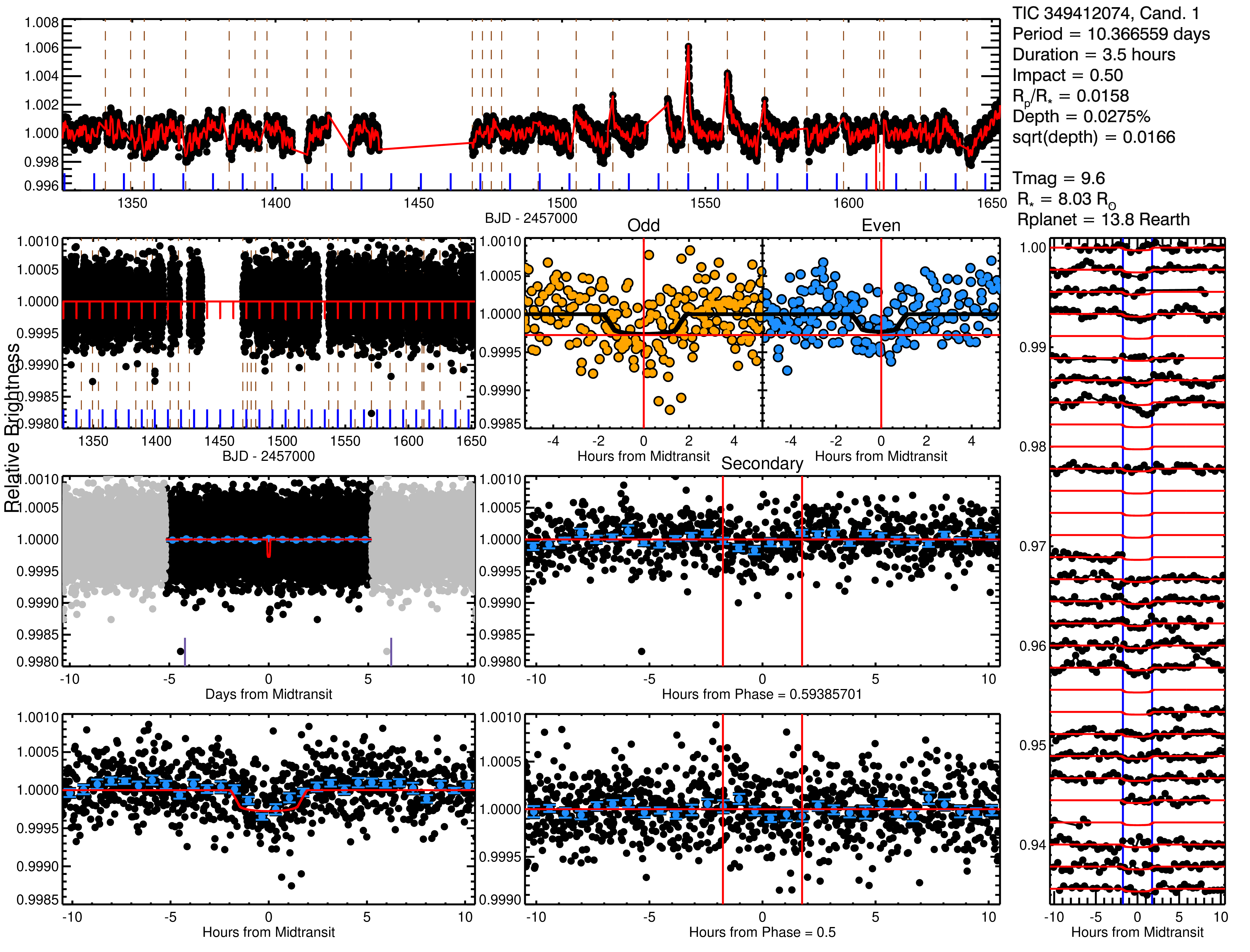}
    \caption{Example of borderline pattern. The very low signal to noise ratio of the transit signal is easily mistaken for a ``J''.}
    \label{fig:hardexamplelowsnr}
\end{figure*}

In our tests, we found a common source of false negatives stemming from patterns with borderline label assessments. The most common being eclipsing binaries which are non-contact but still close enough to resemble the pattern of a contact binary, due to, for example, tidal distortion, hence it is unclear whether the label should be ``E'' or ``B'' (Figure \ref{fig:hardexampleblike}). Other instances of ambiguous patterns are represented by very noisy transits, or transits on a background of high stellar variability, where the distinction between ``E'' and ``J'' is more subtle (Figure \ref{fig:hardexamplelowsnr}).

\begin{figure*}
    \centering
    \includegraphics[width=0.8\textwidth]{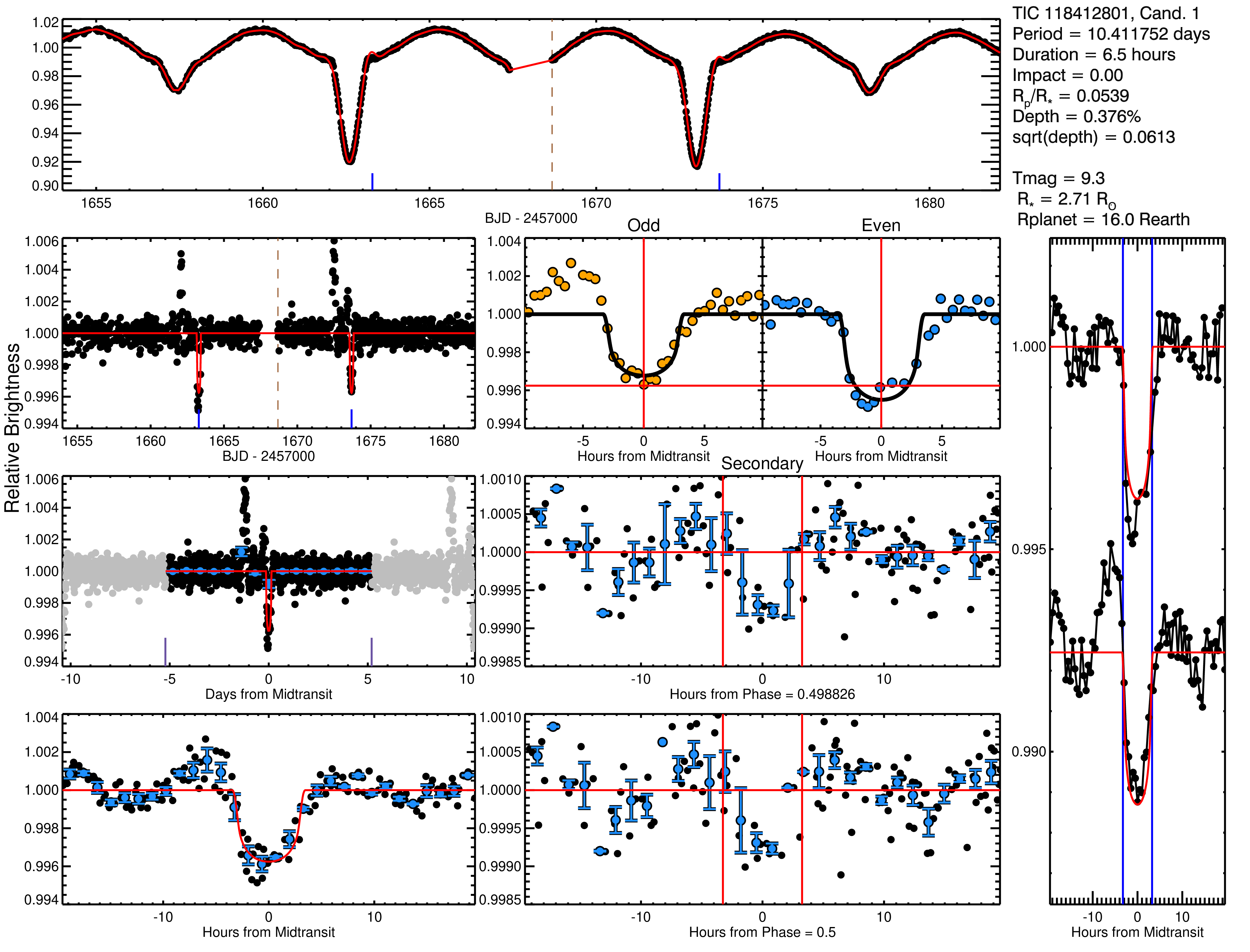}
    \caption{Example of incorrect BLS estimation. Although the phase and period are close, the transit duration is too small, causing the transit to be clipped by the detrending process.}
    \label{fig:hardexamplebadduration}
\end{figure*}

\begin{figure*}
    \centering
    \includegraphics[width=0.8\textwidth]{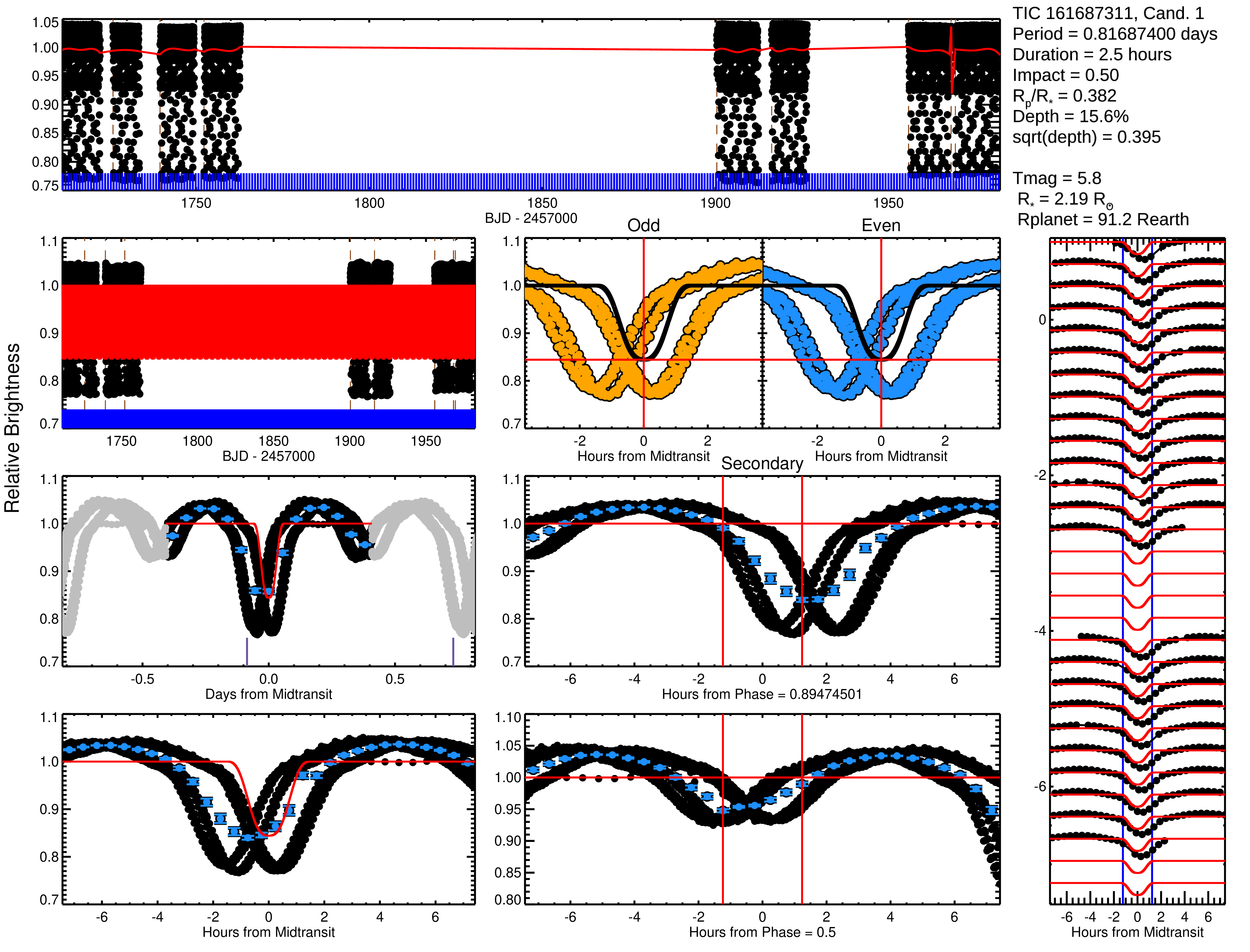}
    \caption{Example of incorrect BLS estimation. The detected period is close, but when the light curve contains a large number of folds, the error compounds and leads to a blurring of the transit view. This is due to QLP searching the light curve with an undersampled BLS frequency grid (necessary due to the computational time needed to run BLS on a large number of targets each sector), as discussed in Kunimoto et al.~(2022, in prep.).}
    \label{fig:hardexampleinexactperiod}
\end{figure*}

One particular element of sensitivity for the neural network is on the correctness of the period and duration values estimated by BLS. Errors in these values can lead to de-trending distortions which can make phase-folded views deviate from a transit-like light curve shape. Examples containing multi-year observations can be particularly sensitive, as even slight variations in the detected period can lead to a blurring of the transit in the phase folded view (Figure \ref{fig:hardexamplebadduration} and  \ref{fig:hardexampleinexactperiod}).

We also note that the phase folding and binning processes are inherently lossy (similar to how compressing an image is a lossy process). While we have not ascertained the impact of such loss of information, it is to be expected that it causes some loss of precision.

\subsection{Comparison to other works}

Our work is largely based on the original TESS \texttt{Astronet-Triage} classifier described by \citet{2019AJ....158...25Y}, which was used for QLP planet candidate triage from Sectors 6 to 33. The following summarizes the major differences in development and implementation between classifiers:

\begin{enumerate}
    \item \texttt{Astronet-Triage} was trained and tested on QLP light curves from only TESS Sectors 1 -- 5, while \texttt{Astronet-Triage-v2} was trained and tested on Sectors 1 -- 39.
    \item \texttt{Astronet-Triage} was developed using 16,516 labeled TCEs (493 planet candidates, 2155 eclipsing binaries, and 13,868 noise/systematic signals), which is roughly two-thirds the size of our labeled set (24,926 TCEs).
    \item \texttt{Astronet-Triage} used labels that were assigned by only a single vetter who visually inspected all TCEs, while 3 -- 5 vetters independently inspected each of the TCEs for \texttt{Astronet-Triage-v2}, and group discussions resolved labeling disagreements. As a result, our labels should be more reliable.
    \item \texttt{Astronet-Triage} only labels signals as either ``planet'' (for all eclipsing signals, including planets and eclipsing binaries) and ``non-planet'' (for other false positives, including pulsating variables, noise and systematics). The five-label model used by \texttt{Astronet-Triage-v2} (E, S, B, J, N) is more flexible and informative.
    \item \texttt{Astronet-Triage} takes the light curves already detrended by QLP, and bins the data into two views: a ``global'' view, showing the full light curve phase diagram, and a ``local'' view, showing a close-up of the transit in the phase diagram. As described in Section \ref{inputrepresentation}, \texttt{Astronet-Triage-v2} creates three sets of detrended light curves from the raw QLP light curve, and generates seven views for each one. In total, \texttt{Astronet-Triage-v2} uses 21 unique views to inform its classification compared to the two used by \texttt{Astronet-Triage}.
\end{enumerate}

These key differences result in improvements to our ability to classify TESS signals in FFI data, as shown Sections \ref{sec:em_performance} and \ref{sec:toi_performance}.

To our knowledge \citet{2019AJ....158...25Y} is the only truly comparable work to ours, in that their source dataset was the TESS Full Frame Images and not the pre-selected targets processed by the SPOC pipeline, and, their goal was to perform triage by identifying all eclipsing signals, rather than separating planet candidates from eclipsing binaries and other false positives. Some other groups have trained and tested neural networks on TESS data from two-minute postage stamps processed by the SPOC pipeline \citep{2020A&A...633A..53O, rao, 2021arXiv211110009V, fiscale, ofman}, and were successful in identifying planet candidates. However, in general, these groups find that the neural network performance is worse on TESS data than a similar network on Kepler data, likely due to TESS's higher \textit{a priori} TCE false positive fraction (due to the larger TESS pixels resulting in more blending) and shorter observational baseline. The false positive rate for FFI targets is likely even higher because a) the targets observed by QLP tend to be fainter than targets observed in postage stamps and blending is more pronounced, and b) the targets observed in the FFIs are more often large, luminous stars like red giants, which are difficult to find planets around, and are photometrically noisy. Therefore, TCEs detected by the QLP likely have an even higher \textit{a priori} false positive probability than TCEs detected by TESS in postage stamp data.  

%Most of these other works aimed to  It is therefore difficult to compare the results of classifications in the TESS FFIs with the postage stamps because of the dramatically different distribution of stellar types and (with the FFI dataset having a much higher rate of giant stars and therefore contaminating signals stellar variability  stellar type and 

%\subsection{Implications for planet detection}\label{impl_planets}

\subsection{Future work}\label{futurework}

\subsubsection{Applications to exoplanet population statistics}\label{sec:population_statistics}

Planet catalogs can be used to characterize exoplanet population statistics through the estimation of occurrence rates. One of the key components of occurrence rate methodologies is a characterization of catalog completeness, reflecting how many planets from the underlying population were missed. A second key component is an understanding of catalog reliability \update{\citep{Bryson2020}}, reflecting how much of the catalog is polluted with false positives. For these reasons, occurrence rate studies require the ability to produce planet catalogs in a fully automated, uniform, and reproducible way, rather than relying on biased manual identification of planet candidates.

\update{NASA's Kepler mission has dominated the past decade of demographics work in large part thanks to the fully automated Kepler Robovetter pipeline, which enabled careful characterization of both completeness and reliability across wide areas of exoplanet parameter space \citep{thompsoncatalog, Christiansen2020}. However,} there is not yet a fully automated TESS planet vetting pipeline. \update{Most previous work has also} focused on 2-minute cadence observations rather than FFIs, \update{which will be less suitable for demographics due to selection biases in 2-minute cadence target lists}. \texttt{Astronet-Triage-v2} is an important step toward uniformly vetted FFI planet catalogs, and it naturally allows for a flexibility in balances between completeness and reliability through the adjustment of prediction thresholds for passing candidates. While the classifier is not yet able to distinguish eclipsing binary false positives from planets (labeling all such signals as ``E'''s), it can be used as a first round of automated and characterizable triage. \update{Future improvements to \texttt{Astronet-Triage-v2} (Section \ref{sec:future_improvements}) are expected to improve the precision and recall, and therefore the completeness and reliability, of any resulting planet catalog.} We have plans to extend \texttt{Astronet-Triage-v2} to be capable of all steps of the vetting process in the future.

\subsubsection{Further improvements to the neural network}\label{sec:future_improvements}

In future work, we suggest a number of additions to further improve the performance of our classifier. 

Over the past few decades, the performance of deep learning classifiers has seen unprecedented success. A large part of this success has been attributed to the increasing size of training datasets. In this work, the number of training examples is relatively low, particularly for the S-labelled class, with a large class-imbalance (see Figure \ref{fig:label_counts}).

A common technique for increasing training datasets, without obtaining new labelled data, is data augmentation. This typically involve applying slight transformations to the training data to produce new data that mimics real observation. Using a combination of a few data augmentation techniques can magnify a training set by several fold and helps reduce over-fitting. In future work, we suggest applying data augmentation methods such as randomly reversing or clipping light curves in time and applying random Gaussian noise to the light curves or scalar features.  We note that these methods were applied in \citet{2018ApJ...869L...7A}, where they showed that the main benefit to data augmentation on exoplanet classification was alleviating model over-fitting, with only a small improvement to model performance. More complex augmentation methods such as fitting a model (e.g.~Gaussian Process, see \citealt{Boone2019}) to the minority class light curves and generating more synthetic data may also help to improve the limited data for some classes.

Since \texttt{Astronet-Triage-v2} is used in production for QLP's monthly planet search, another way to increase our training dataset is to use the existing human vetting work that goes into producing the TOI catalog \citep{guerrero}. As this human vetting is the final step in the TOI release process, there is a high level of quality control in the labels and the signals being vetted are often the most difficult to classify, making them important examples for the model to learn.
%In the model presented in this manuscript, we only use TESS data from Sectors 1-39. In future work, we suggest assessing the possibility of building \textit{continual learning} models (see \citealt{vandeven2018} that can update their trained models as soon as new labelled data becomes available.

% \begin{enumerate}

% \item data augmentation
% \item training data expansion, continuous learning
% \item end-to-end learning
% \item transfer learning
% \item other applications: periodic (phase-folded, e.g. variable stars, EBs),  non-periodic (sliding window?, full view?)

% \end{enumerate}

\section{Conclusion}\label{conclusions}

We have presented \texttt{Astronet-Triage-v2}, a convolutional neural network designed to distinguish astrophysical eclipsing candidates from other phenomena such as stellar variability and instrumental systematics in TESS FFI light curves. The network assigns input signals one of five labels, namely ``E'' for eclipsing signals, ``S'' for single transits or incorrect periods, ``B'' for contact binaries, ``J'' for signals due to noise or systematics, and ``N'' for inconclusive cases. We trained \texttt{Astronet-Triage-v2} using $\sim25000$ signals, which were detected by QLP from TESS Sectors 1 -- 39 and human-labeled through manual review and group discussion. We make this training set available to the community.

\texttt{Astronet-Triage-v2} is the next in a line of \texttt{Astronet} architectures, which were first used for Kepler \citep{shallue2018measuring} and later extended to K2 \cite[\texttt{Astronet-K2};][]{2019AJ....157..169D} and TESS \cite[\texttt{Astronet-Triage};][]{2019AJ....158...25Y}. This iteration features significant improvements over \texttt{Astronet-Triage}, including a larger and more robust training set, an expanded list of possible classifications, and more than ten times the number of unique views used to analyze each signal. As a result, we found \texttt{Astronet-Triage-v2} is more successful at correctly labeling known TOIs across almost all cutoff values, with 86\% recall at a cutoff of 0.215 compared to 82\% recall by \texttt{Astronet-Triage}. When tested on a set of new signals from Sector 33, \texttt{Astronet-Triage-v2} provides better recall of E and S labels than \texttt{Astronet-Triage} for similar (or better) levels of precision, especially for fainter targets. Starting in Sector 34, \texttt{Astronet-Triage-v2} officially replaced \texttt{Astronet-Triage} within QLP.

As both the TESS observing baseline and number of observed stars continue to increase, automated TESS planet vetting tools will become more important. This is especially true of tools tuned for planet searches using FFIs, of which \texttt{Astronet-Triage-v2} is one of the few currently available. While \texttt{Astronet-Triage-v2} is not yet capable of distinguishing between eclipsing binaries and transiting planets, it serves as an effective first round of automated and characterizable triage. We plan to continue to improve and extend the network into a fully automated vetting tool in the future.

\vspace{0.1in}
\section*{Acknowledgements}

% TESS
This paper includes data collected by the TESS mission. Funding for the TESS mission is provided by the NASA's Science Mission Directorate.

% Gaia
This work has made use of data from the European Space Agency (ESA) mission
{\it Gaia} (\url{https://www.cosmos.esa.int/gaia}), processed by the {\it Gaia}
Data Processing and Analysis Consortium (DPAC,
\url{https://www.cosmos.esa.int/web/gaia/dpac/consortium}). Funding for the DPAC
has been provided by national institutions, in particular the institutions
participating in the {\it Gaia} Multilateral Agreement.

% Tansu
This work was supported by an LSSTC Catalyst Fellowship awarded by LSST Corporation to T.D. with funding from the John Templeton Foundation grant ID \#62192.

The \texttt{Astronet-Triage-v2} model was trained and tuned on Google Compute Engine.

%SU, FP, FB, DS, CL, DE, M.Marmier, and M.Mayor acknowledge financial support from the Swiss National Science Foundation (SNSF) in the frame work of the National %Centre for Competence in Research PlanetS. DE acknowledges financial support from the European Research Council (ERC) under the European Union’s Horizon 2020 research and innovation program (project {\sc Four Aces}; grant agreement 724427).
%NN acknowledges partial supported by JSPS KAKENHI Grant Number JP18H01265 and JST PRESTO Grant Number JPMJPR1775.
%We made use of the Python programming language \citep{Rossum1995} 
%and the open-source Python packages
%\textsc{numpy} \citep{vanderWalt2011}, 
%{\scshape scipy} \citep{Jones2001}, 
%{\scshape matplotlib} \citep{Hunter2007}, 
%\textsc{emcee} \citep{Foreman-Mackey2013}, 
%{\scshape george} \citep{Ambikasaran2014}, 
%and
%\textsc{celerite} \citep{Foreman-Mackey2017}.
%{\scshape rebound} \citep{rebound},
%and {\scshape reboundx}. 

%{\scshape corner} \citep{Foreman-Mackey2016}, 
%{\scshape seaborn} (\url{https://seaborn.pydata.org/index.html}),
%===============================================================================

%===============================================================================
\facility{TESS, {\it Gaia}}

\textit{Software:} numpy \citep{np}, matplotlib \citep{plt}, pandas \citep{reback2020pandas, mckinney-proc-scipy-2010}, statsmodels \citep{seabold2010statsmodels}, pydl, astropy \citep{astropy:2013, astropy:2018}, TensorFlow \citep{tf}, Vizier \citep{vizier}, Jupyter \citep{jupyter}

\vspace{0.1in}

\bibliographystyle{apj}
\bibliography{bibliography}

\appendix
% \restartappendixnumbering

\section{Example TCE table}\label{tab:tce_table}
\update{Example TCE table that is passed into \texttt{Astronet-Triage-v2} along-side raw light curve data. All data is available in \citet{evan_tey_2022_7411579}. This table contains information about the signal detected from BLS (epoch, period, duration, depth), information about the host star from TIC 8.2 (TIC ID, $M_*$, $R_*$, TMag). Est $R_*$ is described in Section \ref{sec:scalar_data}, and year describes the year the TCE was detected. MinT and MaxT specify the time range used from the light curve for both detection and input to \texttt{Astronet-Triage-v2}, and Split specifies which dataset (train, val, test) the signal was in. L1-L8 are labels assigned by individuals and Consensus Label is the label agreed upon by the group.}

\begin{table}[h]
\movetabledown=2in
\begin{rotatetable}
\begin{tabular}{rrrrrrrrrrrrrrcccc}
\toprule
   TIC ID &  Period &     Epoch & Duration &  Depth &  TMag & $M_*$ & $R_*$ & Est $R_*$ &  Year &      MinT &      MaxT & Split & Consensus & L1 & L2 & L3 \\
 & (days) & (BTJD) & (days) & (ppm) & & $M_\odot$ & $R_\odot$ & $R_\odot$ & & BTJD & BTJD & & Label &  \\
\midrule
290603338 & 13.6725 & 1629.9162 &    0.212 &    380 &  8.51 &       & 15.62 &     16.00 &     1 & 1624.9693 & 1682.3443 & train &           &  J &  N &  J \\
 32092337 &  1.3228 & 1326.3441 &    0.219 &    110 &  9.82 &  1.41 &  2.43 &      2.43 &     1 & 1325.3226 & 1652.8639 & train &         J &  J &  J &  J \\
278544052 & 11.2664 & 1600.6568 &    0.206 &   2350 & 11.51 &  1.36 &  1.99 &      2.06 &     1 & 1596.7819 & 1682.3445 & train &         J &  J &  J &  J \\
380752037 &  0.5639 & 1657.5494 &    0.069 &   3970 & 11.29 &  1.96 &  2.28 &      2.18 &     1 & 1653.9262 & 1682.3430 & train &         J &  J &  J &  J \\
259863095 & 14.3676 & 1338.8209 &    0.315 &   2660 &  9.56 &  0.82 &  2.71 &      2.50 &     1 & 1325.3233 & 1682.3429 & train &         J &  J &  J &  J \\
272085506 & 42.3842 & 1328.3580 &    0.367 &    890 & 11.39 &       & 14.48 &     15.56 &     1 & 1325.3218 & 1652.8658 & train &         J &  J &  J &  J \\
306897664 &  1.2464 & 1325.8091 &    0.175 &    140 & 10.15 &  3.08 &  3.58 &      3.78 &     1 & 1325.3214 & 1652.8661 & train &           &  J &  N &  B \\
302988499 &  3.1638 & 1493.0635 &    0.085 &  17730 & 11.31 &  1.18 &  1.22 &      1.25 &     1 & 1491.6354 & 1682.3447 & train &         E &  E &  E &  E \\
375033936 & 36.1461 & 1358.3632 &    0.552 &    230 &  9.47 &       & 32.66 &     29.77 &     1 & 1325.3209 & 1682.3446 & train &         J &  J &  J &  J \\
177019341 & 19.7796 & 1326.1799 &    0.296 &    350 & 11.33 &       & 14.13 &     13.35 &     1 & 1325.3215 & 1682.3446 & train &         J &  J &  J &  J \\
 90921748 & 11.3622 & 1660.7688 &    0.163 &   3430 & 10.84 &  2.07 &  2.31 &      2.33 &     1 & 1653.9270 & 1682.3437 & train &         J &  J &  E &  E \\
101738624 &  0.2269 & 1654.1432 &    0.045 &   2010 & 11.44 &  1.72 &  2.91 &      2.72 &     1 & 1653.9256 & 1682.3424 & train &           &  J &  B &  B \\
260415628 & 31.0662 & 1353.6982 &    0.298 &    169 &  8.21 &       & 27.50 &     28.21 &     1 & 1325.3207 & 1682.3445 & train &         J &  J &  J &  J \\
349095939 &  8.6985 & 1331.6535 &    0.152 &    380 & 10.15 &       &  9.53 &     10.57 &     1 & 1325.3211 & 1682.3449 & train &           &  J &  N &  J \\
167344043 &  1.3041 & 1326.4449 &    0.188 &    350 & 11.30 &  0.97 &  1.00 &      1.00 &     1 & 1325.3217 & 1682.3445 &   val &         J &  J &  J &  J \\
339667988 &  3.0673 & 1384.2735 &    0.193 &  34030 &  9.40 &  2.41 &  2.22 &      2.15 &     1 & 1381.7181 & 1682.3452 & train &         E &  E &  E &  E \\
340060373 & 12.9080 & 1393.7434 &    0.157 &   1790 & 11.07 &  1.19 &  1.20 &      1.22 &     1 & 1381.7179 & 1682.3454 &   val &         J &  J &  S &  J \\
340929171 &  1.0873 & 1654.4122 &    0.174 &   2560 & 11.25 &       &       &      1.81 &     1 & 1653.9260 & 1682.3428 & train &           &  B &  J &  J \\
349647610 & 36.1689 & 1359.0938 &    0.587 &    300 &  4.86 &       & 41.18 &     42.82 &     1 & 1325.3211 & 1682.3450 & train &         J &  J &  J &  J \\
340065079 & 16.5423 & 1396.2365 &    0.155 &    570 &  9.60 &       & 10.50 &     10.96 &     1 & 1381.7180 & 1682.3453 & train &           &  J &  N &  J \\
119088593 &  0.3616 & 1654.1666 &    0.157 &  18550 & 10.82 &  1.72 &  1.85 &      1.94 &     1 & 1653.9273 & 1682.3439 & train &         B &  B &  E &  B \\
143769346 &  0.8667 & 1655.7849 &    0.248 &   3720 &  8.99 &  1.65 &  1.65 &      1.66 &     1 & 1653.9264 & 1682.3432 & train &         B &  B &  B &  B \\
320004264 &  1.0463 & 1654.3056 &    0.172 &    570 &  8.19 &  1.34 &  1.25 &      1.25 &     1 & 1653.9262 & 1682.3430 & train &         J &  J &  J &  J \\
 63343395 &  0.3958 & 1657.3129 &    0.039 & 199070 & 10.03 &  1.35 &  1.48 &      1.54 &     1 & 1653.9275 & 1682.3441 & train &         B &  E &  B &  B \\
261543672 & 14.5671 & 1333.9494 &    0.359 &    630 & 10.56 &  1.03 &  1.75 &      1.76 &     1 & 1325.3244 & 1682.3430 & train &         J &  J &  E &  J \\
... & ... & ... & ... & ... & ... & ... & ... & ... & ... & ... & ... & ... & ... & ... & ... & ... & ... \\
\bottomrule
\end{tabular}

\end{rotatetable}
\end{table}

\end{document}